\documentclass{article}
\usepackage[letterpaper,top=2cm,bottom=2cm,left=3cm,right=3cm,marginparwidth=1.75cm]{geometry}
\usepackage[toc,page]{appendix}
\usepackage{amsmath}
\usepackage[most]{tcolorbox}
\usepackage{amssymb} 
\usepackage{amsthm}
\usepackage{float}
\usepackage{physics}
\usepackage{graphicx}
\usepackage{subcaption}
\usepackage{longtable}
\usepackage[version=4]{mhchem}
\usepackage[colorlinks=true, allcolors=blue]{hyperref}

\theoremstyle{plain}  
\newtheorem{prop}{Proposition}
\newtheorem{thm}{Theorem}[section]

\newtheorem{cor}[thm]{Corollary}

\theoremstyle{definition}
\newtheorem{rmk}{Remark}[section]
\newtheorem{dfn}{Definition}[section]

\usepackage{booktabs}

\usepackage{tikz}
\usetikzlibrary{calc,arrows.meta}
\usepackage{xcolor}
\definecolor{dscolor}{RGB}{0, 150, 0}

\usepackage[capitalize]{cleveref}
\crefname{rmk}{Remark}{Remarks}
\crefname{assm}{Assumption}{Assumptions}
\crefname{dfn}{Definition}{Definitions}
\crefname{appsec}{Appendix}{Appendices}
\Crefname{appsec}{Appendix}{Appendices}

\crefname{prop}{Proposition}{Propositions}
\crefname{thm}{Theorem}{Theorems}
\crefname{lem}{Lemma}{Lemmas}
\crefname{clm}{Claim}{Claims}
\crefname{cor}{Corollary}{Corollaries}
\crefname{fact}{Fact}{Facts}

\newcommand{\rchi}{\mathord{\raisebox{0.15ex}{\(\chi\)}}}

\title{Quantum Loading of Charge and Spin-Adapted MPS}
\author{
Matthew R. Hennefarth\thanks{\texttt{mhennefarth@psiquantum.com}}, \quad
Lesley Gover\thanks{\texttt{lgover@psiquantum.com}},\quad
Carys Harvey\thanks{\texttt{charvey@psiquantum.com}},\quad
Zheting Jin,
\\[0.35em]
Ryan Levy,\quad
Sukin Sim, \quad
Modjtaba Shokrian Zini
\\[0.7em]
\textit{PsiQuantum, 700 Hansen Way, Palo Alto, CA 94304, USA}
}

\date{}

\begin{document}
\maketitle

\begin{abstract}
Matrix product states (MPS) are powerful classical representations of correlated quantum states, but loading a large bond dimension MPS can substantially increase the cost in fault-tolerant quantum algorithms. We develop a circuit for preparing a U(1)~$\times$~SU(2)-symmetric MPS directly from spin-adapted reduced tensors. Using Householder--LKS synthesis, the circuit prepares each symmetry-allowed reduced block and applies Clebsch--Gordan coefficients to recover the magnetic degrees of freedom, avoiding explicit loading of the corresponding unfolded dense tensors. The reduced-tensor data loading is dominant in cost. We benchmark $4 \times 4$ and $4 \times 6$ Fermi--Hubbard states and active-space models of \ce{[Fe2S2]^2-}, \ce{[Fe4S4]^2-}, and low- and high-spin cytochrome P450. At 15-bit rotation precision, the molecular benchmarks show $\sim 14\times$ to $54\times$ Toffoli reductions, $\sim 20\times$ to $87\times$ active volume reduction, and $\sim 33\times$ to $140\times$ volume (Toffoli $\times$ qubits) reduction at the QROM-optimal point, relative to dense MPS loading. Fermi--Hubbard speedups increase with bond dimension and track the inverse tensor density, with modest qubit overhead. However, while we discuss generalization of our algorithm to reduce cost given further symmetry or using more performant unitary synthesis methods, extrapolation to larger lattices suggests that DMRG-state loading may exceed downstream phase-estimation costs, motivating co-design of state fidelity, compression, and the initial-state ansatz.
\end{abstract}

\tableofcontents
\section{Introduction}

Fault-tolerant quantum algorithms offer a path for simulations of chemistry and condensed matter systems with accuracy that goes beyond state-of-the-art classical methodologies.
Many of these algorithms have theoretical speedups but are inherently tied to the quality of the initial state loaded onto the quantum computer. While such algorithms, such as quantum phase estimation (QPE), are thought to have exponential advantage over classical methodologies assuming efficient initial-state preparation, the orthogonality catastrophe hypothesis suggests preparing a suitable initial state could be exponentially hard yielding an overall quantum algorithm with polynomial advantage in the best-case scenario \cite{lee2023evaluating}.
The initial-state preparation is often an overlooked cost, despite being vital for achieving quantum advantage in simulating physical and chemical systems.
While ignoring this cost may be valid for estimating particular resources such as gate complexity, initial-state preparation must be considered to fully realize the algorithm and be able to estimate the total time-to-solution.
Previous studies have proposed various initial-state preparation algorithms including adiabatic protocols \cite{aspuruguzik2005simulated}, filtering-based methods \cite{gilyen2019quantum, lin2020optimal, lin2020near, motta2020determining, fomichev2024initial, wang2022state, keen2021quantum, filip2025beyond}, and dissipative dynamics \cite{li2025dissipative, li2026dissipative}.

Combined with the density matrix renormalization group (DMRG) algorithm, matrix product states (MPS) have become the state-of-the-art approach for studying strongly correlated one-dimensional systems. These techniques routinely achieve near-exact ground-state energies and correlation functions for systems beyond the reach of exact diagonalization, and have become a standard tool in condensed matter physics, quantum chemistry, and materials science. Their ability to compactly represent the low-energy eigenstates of chemical and condensed matter Hamiltonians also makes MPSs attractive initial states for fault-tolerant quantum algorithms \cite{fomichev2024initial, berry2025rapid, rupprecht2026faster}. Once prepared on the quantum computer, these states can be evolved under the full Hamiltonian without the need to maintain an explicit tensor-network representation.

Tensor network techniques owe their success to the fact that quantum states arising in physical systems are often highly structured and occupy only a small corner of the exponentially large Hilbert space. Utilizing this fact, a wave function can be represented using a number of parameters that grows only polynomially with system size. To further reduce the overhead of the polynomial growth, many quantum systems of interest are also invariant under spin rotations, including non-relativistic electronic-structure Hamiltonians that describe molecules and materials as well as model Hamiltonians in the absence of magnetic interactions, such as the Fermi--Hubbard model \cite{Hubbard1963}. Such Hamiltonians have eigenstates that may be labeled by a number of symmetries, including the SU(2) spin rotation symmetry often accompanied by a U(1) charge symmetry such as total particle number. Some models have an additional symmetry at a particular electronic filling (e.g. half filling $n_{\text{filling}}=1$) that has a SU(2) charge symmetry giving the overall symmetry of SU(2)~$\times$~SU(2)~$=$~SO(4) \cite{Mielke2015}. Exploiting these symmetries during initial-state preparation restricts the state to the correct symmetry sector, avoiding leakage into symmetry-incompatible sectors and further reducing the amount of information one has to load and store on the quantum computer (see \cref{sec:qpe_emulation}).
Additionally, this has been shown to have a significant improvement to the classical storage efficiency and runtime costs.
By adding just a single U(1) symmetry to the MPS tensors, they gain a 20\% sparsity structure, with a second additional U(1) symmetry reducing the sparsity to 5\% \cite{levy2020distributed}.
Symmetry restrictions can also remove irrelevant states during calculations, tightly targeting the correct state \cite{McCulloch2002,Sharma2012}.
Recently Ref.~\cite{rupprecht2026faster} exploits the U(1) symmetry in loading an MPS to achieve a $10$ to $30$ times improvement in Toffoli count for molecular systems that are often benchmarks for quantum simulation applications.

In this work, we compile a quantum circuit for loading a charge and spin-adapted U(1)~$\times$~SU(2)-symmetric MPS based on the standard Householder-LKS unitary synthesis method. We may at times drop the U(1) symmetry for less cluttering. 
We provide quantum resource estimates (QREs) in terms of Toffoli, active volume (\cite{litinski2022active}), and qubit counts for the Fermi--Hubbard model and some molecular systems including comparisons to the equivalent loading of a dense MPS, all using verified circuits implemented in Workbench \cite{psiquantum_qdk}. The main positive conclusion is that the structure of the SU(2) tensor can be encoded with minimal overhead and thus we can lower the gate count of the MPS loading by a factor that grows with the sparsity. 
However, the trend of costs is worrying for the lattice Hamiltonians. 
These models are typically interesting at larger sizes due to highly performant classical methods, and thus for early-generation fault-tolerant applications we have less space for qubit-gate tradeoffs. 
On the other hand, the larger the model gets the much larger the bond dimension needs to be, and this means a cost that is orders of magnitude higher than that of the downstream task such as the QPE \cite{kan2024resource}. Our results are also compared to the recent \cite{rupprecht2026faster} for several molecular systems, where we see that the gate-qubit tradeoff in QROMs has higher impact on the dense MPS loading than on the SU(2) version precisely due to the latter's block-sparse structure. 

On the more positive note for future works, we show how efficient unitary synthesis methods such as those in \cite{rupprecht2026faster, berry2025rapid} are orthogonal to this work, and can be employed for further speedup, to attain low gate and qubit count. We also note how other non-DMRG alternatives, such as Hartree-Fock and its variants, could provide potentially cheaper initial state loading modality for lattice systems.

\section{SU(2)-symmetric MPS}

\subsection{Basics}

Consider a one-dimensional quantum system consisting of $N$ lattice sites. In an MPS representation, the many-body wave function is factorized into a sequence of local tensors.
\begin{equation}
    \ket{\Psi} = \sum_{\{P^i\}} \sum_{\{\rchi_i\}} \left( \prod_{i=1}^{N} (T^{[i]})^{\rchi_i}_{\rchi_{i-1},P^i} \right) \ket{P^1 P^2 \cdots P^N}
\end{equation}
Associated with each physical site, $i$, is a tensor, $T^{[i]}$. This tensor is composed of a physical index, $0 \leq P^i < d$, and two virtual indices,
$\rchi_{i-1},\rchi_{i}$. The notation $\rchi$ will be abused throughout the paper, and it will be clear from context where it refers to an index or to the bond itself. The physical index enumerates a $d$-dimensional Hilbert space encoding local degrees of freedom. The virtual indices connect neighboring tensors and enumerate many-body Schmidt states encoding the entanglement structure. The dimensions of $\Bqty{\rchi_i}_{i=0}^{N}$, are referred to as bond dimensions where the maximum value bounds the Schmidt rank across the bond, with the first and last being the vacuum and fully occupied bonds. 
$\rchi_{\mathrm{max}} = \max_i \rchi_i$ is commonly referred to as the bond dimension of the MPS.
Using these local tensors, we can represent a wave function with $O\pqty{Nd \rchi^2}$ variables, i.e., polynomial in the number of sites. More high-level illustration of the tensor is available in \cref{appsec:ill_block}. 

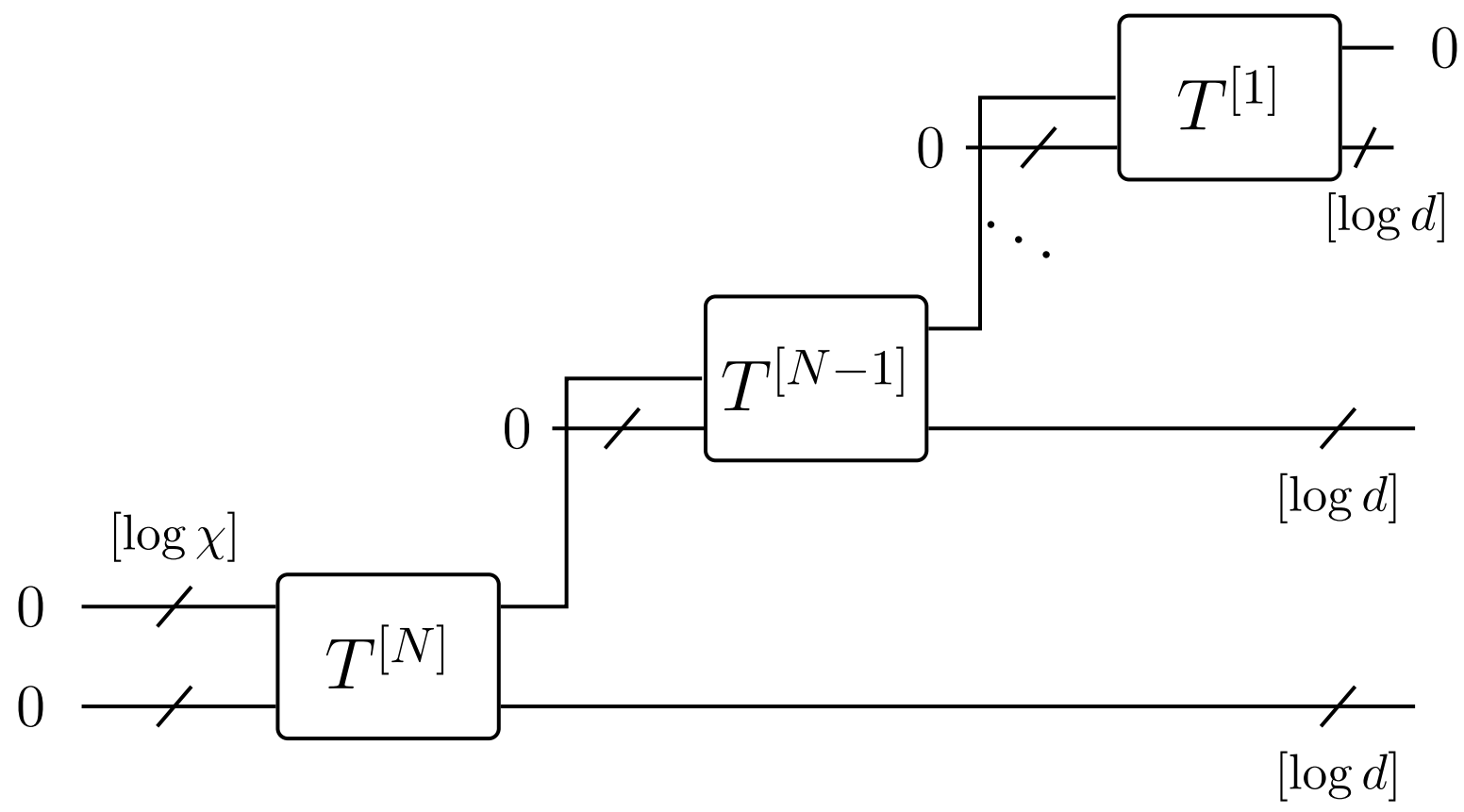
\begin{figure}[t]
\centering

\begin{tikzpicture}[
  x=1cm,
  y=1cm,
  wire/.style={line width=0.75pt},
  gate/.style={
      draw,
      rounded corners=2pt,
      line width=0.75pt,
      minimum width=1.55cm,
      minimum height=1.15cm,
      inner sep=1pt,
      font=\Large
  },
  lab/.style={font=\large},
  smalllab/.style={font=\normalsize}
]

\node[gate] (UN) at (0,0) {$T^{[N]}$};

\draw[wire] (-2.15,0.35) -- (UN.west |- 0,0.35);
\node[lab,anchor=east] at (-2.28,0.35) {$0$};

\draw[wire] (-1.62,0.21) -- (-1.38,0.49);
\node[smalllab,anchor=south] at (-1.50,0.54)
    {$[\log\chi]$};

\draw[wire] (-2.15,-0.35) -- (UN.west |- 0,-0.35);
\node[lab,anchor=east] at (-2.28,-0.35) {$0$};

\draw[wire] (-1.62,-0.49) -- (-1.38,-0.21);

\draw[wire]
    (UN.east |- 0,-0.35) -- (7.20,-0.35);

\draw[wire]
    (6.54,-0.49) -- (6.78,-0.21);

\node[smalllab,anchor=north]
    at (6.66,-0.54) {$[\log d]$};

\draw[wire]
    (UN.east |- 0,0.35)
    -- (1.25,0.35)
    -- (1.25,1.95)
    -- (2.20,1.95);

\node[gate] (UNm) at (3.00,1.95) {$T^{[N-1]}$};

\draw[wire]
    (1.15,1.60) -- (UNm.west |- 0,1.60);

\node[lab,anchor=east]
    at (1.13,1.60) {$0$};

\draw[wire]
    (1.52,1.46) -- (1.76,1.74);

\draw[wire]
    (UNm.east |- 0,1.60) -- (7.20,1.60);

\draw[wire]
    (6.54,1.46) -- (6.78,1.74);

\node[smalllab,anchor=north]
    at (6.66,1.41) {$[\log d]$};

\draw[wire]
    (UNm.east |- 0,2.30)
    -- (4.15,2.30)
    -- (4.15,3.92)
    -- (5.10,3.92);

\node[font=\Large] at (4.42,3.03) {$\ddots$};

\node[gate] (Uone) at (5.90,3.92) {$T^{[1]}$};

\draw[wire]
    (4.05,3.57) -- (Uone.west |- 0,3.57);

\node[lab,anchor=east]
    at (4.03,3.57) {$0$};

\draw[wire]
    (4.44,3.43) -- (4.68,3.71);

\draw[wire]
    (Uone.east |- 0,4.27) -- (7.05,4.27);

\node[lab,anchor=west]
    at (7.18,4.27) {$0$};

\draw[wire]
    (Uone.east |- 0,3.57) -- (7.05,3.57);

\draw[wire]
    (6.78,3.43) -- (6.92,3.71);

\node[smalllab,anchor=north]
    at (7.00,3.38) {$[\log d]$};

\end{tikzpicture}

\caption{\label{fig:circuit} Translating the MPS to a quantum circuit with qubits corresponding to the virtual indices shared between unitaries. The MPS is left-canonical and we go from the right to left in loading the isometries defined by the tensors.}

\end{figure}

To execute an MPS as a quantum circuit, each tensor must be embedded into a unitary operator acting on qubits. 
In the canonical form, each $T^{[i]}$ is an isometry rather than unitary in general; although this is fine for our algorithm, some approaches need to pad the isometry into a unitary \cite{rupprecht2026faster}. The virtual indices are represented by a register of $\lceil \log_2\pqty{\rchi_{\max}}\rceil$ qubits. This virtual register is shared between all tensors in the circuit construction, as shown in \cref{fig:circuit}, and is propagated sequentially through the chain.

For a dense MPS ansatz, the virtual indices are simply integer labels ranging from zero to the maximum bond dimension. An alternative approach is to exploit global symmetries to organize the virtual space into sectors labeled by conserved quantum numbers. Rather than treating each virtual index as a single integer, the index is decomposed into a collection of quantum numbers and degeneracy labels. Specifically, we will consider the U(1) symmetry associated with particle-number conservation and the SU(2) symmetry associated with spin-rotation symmetry. States are therefore classified by the quantum numbers $(n, S)$, where $n$ is the particle number and $S$ is the total spin.

The irreducible representations of SU(2) are the spin multiplets consisting of the $2S + 1$ states, $-S \leq m_S \leq S$. Equivalently, the vector space $V_S$ spanned by these states remains invariant under the action of all SU(2) operators and contains no smaller invariant subspace. The virtual bond space (compressed Schmidt space) can then be decomposed into symmetry sectors labeled by the conserved quantum numbers $\pqty{n, S}$. Each sector may appear multiple times. We call the number of times a sector occurs the multiplicity, $C_{n, S}$ which determines the dimension of the corresponding degeneracy space, $D_{n, S}$.
The full virtual space takes the form
\begin{equation}
    \bigoplus_{n,S} D_{n,S} \otimes V_S.
\end{equation}
Note that for a dense MPS, the truncation error and variational accuracy are controlled by the maximum bond dimension, $\rchi_\mathrm{max}$. For an SU(2) symmetric MPS, the analogous variational resource is the set of retained symmetry sectors, $\Bqty{\pqty{n, S}}$, and their degeneracies, $C_{n,S}$, which collectively define an effective bond dimension. In practice, the selection of these hyperparameters are determined via DMRG.

We will work in second-quantization with the physical leg represented by two qubits, one representing the spin-up occupation and one spin-down. We consider $S\in\Bqty{0, \frac{1}{2}}$ and correspondingly $n\in[0,1,2]$ such that we can encode all possible states for a given site $P$ via two qubits as
\begin{align}
        \ket{00}: & \hspace{0.5cm} \pqty{n_{P},S_{P}, m_P} = \pqty{0,0,0} \\
        \ket{10}: & \hspace{0.5cm} \pqty{n_{P},S_{P}, m_P} = \pqty{1,\frac{1}{2},\frac{1}{2}}
        \\
        \ket{01}: & \hspace{0.5cm} \pqty{n_{P},S_{P}, m_P} = \pqty{1,\frac{1}{2}, -\frac{1}{2}}
        \\
        \ket{11}: & \hspace{0.5cm} \pqty{n_{P},S_{P}, m_P} = \pqty{2,0,0},
\end{align}
and the physical leg can be described by the dictionary,
\begin{align}\label{eq:P_i}
    P = \{ (0,0) : \text{span}\{\ket{00}\} , (1,\frac{1}{2}) : \text{span}\{ \ket{10},\ket{01}\}, (2,0): \text{span}\{\ket{11} \} \ \}.
\end{align}

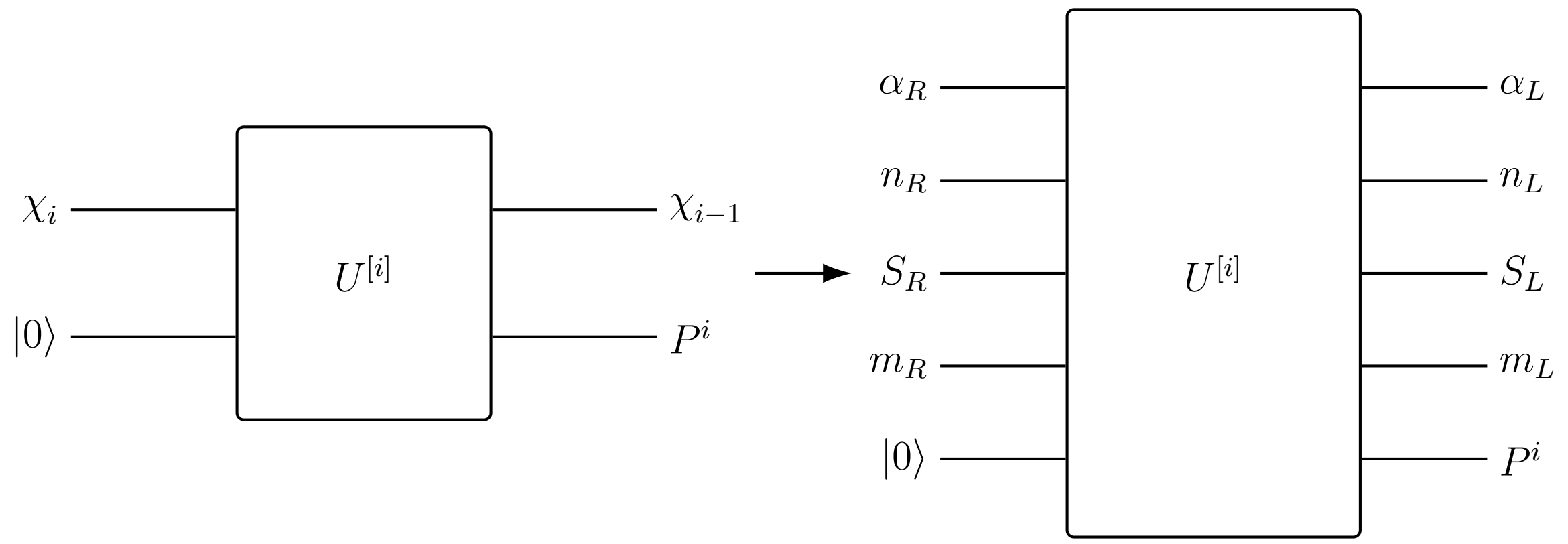
\begin{figure}[t]
\centering
\resizebox{\textwidth}{!}{%
\begin{tikzpicture}[
    font=\large,
    box/.style={
        draw,
        rounded corners=2pt,
        line width=0.8pt,
        minimum width=2.6cm,
        minimum height=3.0cm
    },
    bigbox/.style={
        draw,
        rounded corners=2pt,
        line width=0.8pt,
        minimum width=3.0cm,
        minimum height=5.4cm
    },
    wire/.style={line width=0.8pt},
    expandarrow/.style={
        -{Latex[length=3mm,width=2mm]},
        line width=0.8pt
    }
]

\node[box] (U1) at (0,0) {$U^{[i]}$};

\draw[wire] (-3.0, 0.65) -- (U1.west |- 0,0.65);
\node[left] at (-3.0,0.65) {$\rchi_i$};

\draw[wire] (-3.0,-0.65) -- (U1.west |- 0,-0.65);
\node[left] at (-3.0,-0.65) {$\lvert 0\rangle$};

\draw[wire] (U1.east |- 0,0.65) -- (3.0,0.65);
\node[right] at (3.0,0.65) {$\rchi_{i-1}$};

\draw[wire] (U1.east |- 0,-0.65) -- (3.0,-0.65);
\node[right] at (3.0,-0.65) {$P^i$};

\draw[expandarrow] (4.0,0) -- (5.0,0);

\node[bigbox] (U2) at (8.7,0) {$U^{[i]}$};

\def\ya{1.90}
\def\yn{0.95}
\def\ys{0.00}
\def\ym{-0.95}
\def\yz{-1.90}

\draw[wire] (5.9,\ya) -- (U2.west |- 0,\ya);
\node[left] at (5.9,\ya) {$\alpha_R$};

\draw[wire] (5.9,\yn) -- (U2.west |- 0,\yn);
\node[left] at (5.9,\yn) {$n_R$};

\draw[wire] (5.9,\ys) -- (U2.west |- 0,\ys);
\node[left] at (5.9,\ys) {$S_R$};

\draw[wire] (5.9,\ym) -- (U2.west |- 0,\ym);
\node[left] at (5.9,\ym) {$m_R$};

\draw[wire] (5.9,\yz) -- (U2.west |- 0,\yz);
\node[left] at (5.9,\yz) {$\lvert0\rangle$};

\draw[wire] (U2.east |- 0,\ya) -- (11.5,\ya);
\node[right] at (11.5,\ya) {$\alpha_L$};

\draw[wire] (U2.east |- 0,\yn) -- (11.5,\yn);
\node[right] at (11.5,\yn) {$n_L$};

\draw[wire] (U2.east |- 0,\ys) -- (11.5,\ys);
\node[right] at (11.5,\ys) {$S_L$};

\draw[wire] (U2.east |- 0,\ym) -- (11.5,\ym);
\node[right] at (11.5,\ym) {$m_L$};

\draw[wire] (U2.east |- 0,\yz) -- (11.5,\yz);
\node[right] at (11.5,\yz) {$P^i$};

\end{tikzpicture}
}
\caption{
Unitary expansion of the SU(2)-symmetric tensor at site $i$.
The right virtual index basis state
$\ket{\rchi_i=(\alpha_R,n_R,S_R,m_R)}$
labels the input, while the left virtual index basis state
$\ket{\rchi_{i-1}=(\alpha_L,n_L,S_L,m_L)}$
and physical index $P^i$ label the output.
}
\label{fig:su2_unitary}
\end{figure}

The virtual legs are represented by
\begin{align}
    \rchi = \{ (\alpha,n,S) : \text{span}\{ \ket{\alpha,n,S,m_{S}}_{\rchi} | m_S \in [-S, S] ; \  1 \le \alpha \le C_{n,S} \}\},
\end{align}
where, as we have seen, $C_{n,S}$ is the multiplicity of $\pqty{n,S}$ and $\alpha$ enumerates the copies. When it is clear which tensor we are referring to, we will use the subscript $L,R$ to denote the left and right virtual indices of the tensor as shown in \cref{fig:su2_unitary}. The first vacuum bond is always given by $\rchi_0 = \{(1,0,0)\}$ and the target particle-number and total-spin boundary by $\rchi_N = \{(1,f,S)\}$ where $f,S$ are the particle number and spin target.

\subsection{Reduced Tensor and Clebsch--Gordan Coefficients}\label{sec:cg_restraints}

By the Wigner--Eckart theorem, an SU(2)-symmetric tensor factorizes into a reduced tensor, $A$, carrying the variational degrees of freedom and a symmetry tensor, $CG$, composed of Clebsch--Gordan coefficients
\begin{equation}
    T^{[i]} = A^{(\alpha_R,n_R,S_R)}_{(\alpha_L,n_L,S_L),(n_P, S_P)} \otimes CG^{S_R,m_R}_{S_L,m_L;\,S_P,m_P}.
    \label{eq:TAC}
\end{equation}
The reduced tensor acts on the reduced basis $\pqty{\alpha,n,S}$ and contains all variational parameters. It specifies which symmetry sectors are coupled, subject to the conservation laws of the symmetry. In particular, the physical state $P$, carrying quantum numbers $\pqty{n_P,S_P}$, can only connect virtual sectors $\pqty{n_L,S_L}$ and $\pqty{n_R,S_R}$ that satisfy the allowed particle-number and spin-fusion rules,
\begin{subequations}
    \begin{gather}
        n_L^{i-1}+n_P^i = n_R^i \\ 
        S_L^{i-1} + S_P^{i} + S_R^{i} \in \mathbb{Z}^{\geq 0} \\
        \abs{S_L^{i-1} - S_P^{i}} \le S_R^{i} \le S_L^{i-1} + S_P^{i}
    \end{gather}
\end{subequations}

The Clebsch--Gordan tensor contains no variational freedom and is completely determined by symmetry. It specifies how the magnetic quantum numbers combine within the coupled spin sectors with amplitudes given by the corresponding Clebsch--Gordan coefficients, $\braket{S_L,m_L; S_P,m_P}{S_R,m_R}$. The Clebsch--Gordan coefficients, $CG_{S_Lm_L,S_Pm_P}^{S_R,m_R}$, are non-zero only if,
\begin{equation}
    m_L+m_P=m_R
\end{equation}

For spin-1/2 fermionic sites, we only need to consider a small set of subcases based on the physical spin sectors:
\begin{itemize}
    \item $S_P = 0$:
        \begin{itemize}
            \item $m_P=0$: $m_L=m_R$ with trivial case $S_L = S_R$, $CG_{S_Lm_L,S_Pm_P}^{S_R,m_R}= \delta_{m_L, m_R}$.
        \end{itemize}
    \item $ S_P = 1/2$:
        \begin{itemize}
            \item $m_P\in\{-1/2, 1/2\}$:
            $m_L \in \{m_R-1/2, m_R+1/2\}$ and we now have two cases,
                \begin{itemize}
        \item $S_L = S_R - 1/2$, in which case:
        \begin{align}\label{eq:CGcoeffS_Rminushalf}
            CG_{S_L(m_R-1/2),S_P\frac{1}{2}}^{S_Rm_R} = \sqrt{\frac{S_L+m_R+\frac{1}{2}}{2S_L+1}} \ , \ CG_{S_L(m_R+1/2),S_P\frac{-1}{2}}^{S_Rm_R} = \sqrt{\frac{S_L-m_R+\frac{1}{2}}{2S_L+1}} \ .
        \end{align}
        \item $S_L=S_R+1/2$, in which case:
        \begin{align}\label{eq:CGcoeffS_Rplushalf}
            CG_{S_L(m_R-1/2),S_P\frac{1}{2}}^{S_Rm_R} = -\sqrt{\frac{S_L-m_R+\frac{1}{2}}{2S_L+1}} \ , \ CG_{S_L(m_R+1/2),S_P\frac{-1}{2}}^{S_Rm_R} = \sqrt{\frac{S_L+m_R+\frac{1}{2}}{2S_L+1}} \ .
        \end{align}
    \end{itemize}
                
            \end{itemize}
\end{itemize}

Notice how in all cases, the coefficients describe a real unit vector (of dimension one or two), and in general, for a fixed $S_L,S_P,m_R$, we have
\begin{align}\label{eq:CG_isometry}
    \sum_{\substack{m_L+m_P = m_R \\ -S_L \leq m_L \leq S_L \\ -S_P \leq m_P \leq S_P}} CG_{S_Lm_L;S_Pm_P}^{S_Rm_R} CG_{S_Lm_L;S_Pm_P}^{S_R'm_R} = \delta_{S_R,S_R'} \ .
\end{align}

Thus, the reduced tensor determines which irreducible representations may couple, while the Clebsch--Gordan coefficients determine how the basis states within those irreducible representations are combined. This decomposition reflects the block-sparse structure imposed by SU(2) symmetry where many tensor elements are constrained to be exactly zero, substantially reducing the number of variational parameters relative to a dense MPS.

It is useful to think of the reduced tensor at each site as a collection of the blocks defined by the symmetry sector $(n_R, S_R) \in \rchi_i$,
\begin{equation}
    A^{(\alpha_R, n_R, S_R)}_{(\alpha_L, n_L, S_L), (n_P, S_P)}  
    = \{ (B_{n_R, S_R})^{\alpha_R}_{(\alpha_L,n_L,S_L),(n_P,S_P)}\} 
\end{equation}
where $B_{n_R, S_R}$ is in general a rectangular matrix with rows $\{(\alpha_L, n_L, S_L),(n_P, S_P)\} \in \rchi_{i-1}\times P^i$ and columns $\{(\alpha_R, n_R, S_R)\} \in {\rchi_i}$. Each block will be dense with rows and columns satisfying the Clebsch--Gordan fusion rules. Moreover, each block with be itself an isometry as defined by
\begin{align}\label{eq:A_isometry}
    \sum_{(\alpha_L,n_L,S_L),(n_P,S_P)}(A^{[i]})^{(\alpha_R,n_R,S_R)}_{(\alpha_L,n_L,S_L),(n_P,S_P)}  \overline{(A^{[i]})^{(\alpha_R',n_R,S_R)}_{(\alpha_L,n_L,S_L),(n_P,S_P)}} = \delta_{\alpha_R,\alpha_R'}.
\end{align}

\subsection{Definitions of Dimensions}

We define the size of the virtual qubit register to be
\begin{align}
        \text{size}(\rchi) = \lceil\log_2(\max_{(n_i,S_i) \in \rchi_i,i} C_{n_i,S_i})\rceil+\lceil\log_2(\max_{n_i \in \rchi_i,i} (n_i+1))\rceil+2\lceil\log_2(\max_{S_i \in\rchi_i,i}  2S_i+1)\rceil.
\end{align}
as we need to represent the multiplicity, the particle number starting from zero to $n_i$, and the half-integers spin and magnetic momenta. We define the (reduced) bond dimension of bond $\rchi_i$ via the multiplicities as
\begin{align}
    \dim \rchi_i =  \sum_{(n,S)\in \rchi_i} C_{n,S}^{[i]},
\end{align}
and the \textit{unfolded} bond dimension which includes the magnetic momenta as
\begin{align}
    \dim_{m} \rchi_i := \sum_{(n,S)\in \rchi_i} C_{n,S}^{[i]}(2S+1).
\end{align}

We define the dimension of the tensor $A^{[i]}$ as the sum of the dimension of its blocks
\begin{align}
    \dim_{\mathrm{SU}(2)} A^{[i]} := \sum_{(n,S)\in \rchi_i} \dim(B_{n,S}^{[i]}).
\end{align}
The above is upper bounded by
\begin{align}\label{eq:dimmaxchi_i}
    \dim^{\le}_{\mathrm{SU}(2)} A^{[i]} := \sum_{(n,S)\in \rchi_i} C_{n,S}^{[i]} M_{n,S}^{[i]}
\end{align}
where $M_{n,S} = C_{n-1,S-1/2}^{[i-1]}+C_{n-1,S+1/2}^{[i-1]}+C_{n-2,S}^{[i-1]}+C_{n,S}^{[i-1]}$ is the sum of the multiplicities of all possible $n_L,S_L$ in $\rchi_{i-1}$ that \textit{could} fuse into $n_R=n,S_R=S$.

\section{Circuit Compilation}

To `unfold' the SU(2) MPS we choose a basis $\bigotimes_{j=1}^N(n_{P^j}, S_{P^j}, m_{P^j})$ and a set of labels,
$\kappa:=\{\rchi_i\}_{i=0}^N$, 
from $0$ (left) to $N$ (right) from each bond $\rchi_i$ such that the fusion rules are satisfied. The amplitude in this basis is given by,
\begin{equation}
    \sum_{\kappa}\prod_{i=1}^N (T^{[i]})^{\rchi_i}_{\rchi_{i-1}, P^i} = \sum_{\kappa}\prod_{i=1}^N(A^{[i]})^{(\alpha_R, n_R, S_R)}_{(\alpha_L, n_L, S_L), (n_{P^i}, S_{P^i})}CG^{S_R,m_R}_{S_L, m_L;S_{P^i}, m_{P^i}}.
\end{equation}
We may drop the redundant superscript $i$ index whenever it is clear from the context. Our goal is to compile a circuit representation of the merged tensor, $T^{[i]}$, defined by,
\begin{equation}\label{eq:final_t_column}
\begin{split}
    \ket{(T^{[i]})^{\rchi_i}} &= \sum_{\rchi_{i-1}, P^i} (T^{[i]})^{\rchi_i}_{\rchi_{i-1}, P^i}\ket{\rchi_{i-1}}_{\rchi} \ket{P^i}_i \\
    &= \sum_{\rchi_{i-1}, P^i} (A^{[i]})^{(\alpha_R, n_R, S_R)}_{(\alpha_L, n_L, S_L), (n_P, S_P)} CG^{S_R, m_R}_{S_L, m_L; S_P, m_P} \ket{\alpha_L, n_L, S_L, m_L}_{\rchi} \ket{n_P, S_P, m_P}_i \ .
\end{split}
\end{equation}

\subsection{Background on Householder-LKS unitary synthesis}

Each unitary tensor can be synthesized column by column via a sequence of state preparations and reflections \cite{low2024trading}. 
\begin{figure}[h!]
    \centering
    \includegraphics[width=1.0\columnwidth]{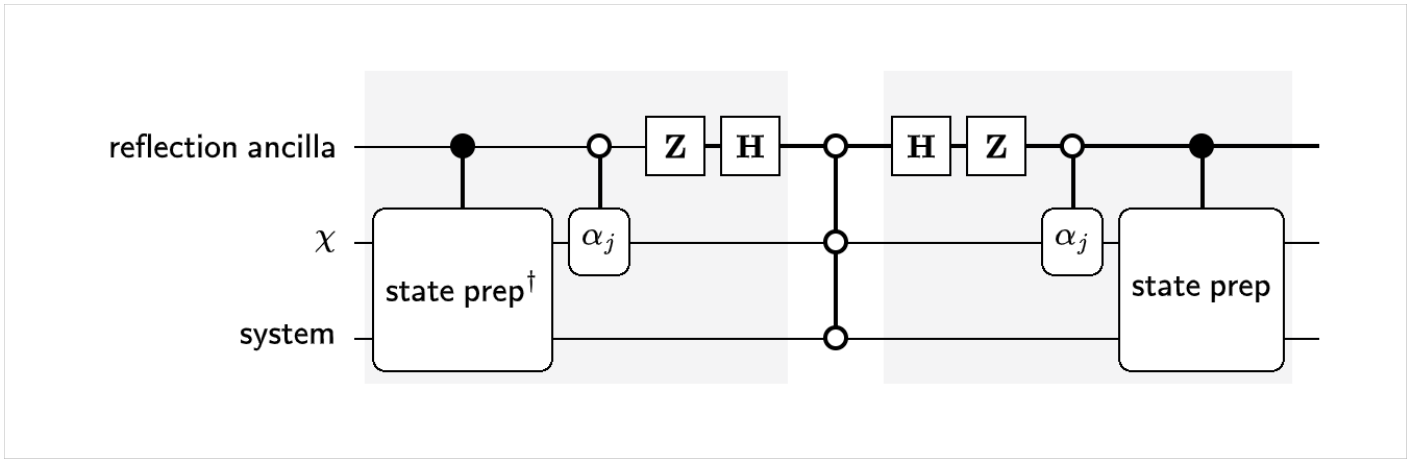}
    \caption{\label{fig:mps_loading}Loading an isometry via Householder decomposition. A reflection ancilla X gate follows to reset the ancilla to 0.}
\end{figure}
To achieve this for a unitary $V$, we introduce a reflection ancilla and define a product of reflections over all columns $v_j$, given by,
\begin{equation}
    \widetilde{V} = X_{\text{reflect}}\prod_{j}(1-2\ketbra{w_j}{w_j})
\end{equation}
where,
\begin{equation}
    \ket{w_j} = \frac{1}{\sqrt{2}}(\ket{0}_\mathrm{reflect}\otimes\ket{j, 0} -  \ket{1}_\mathrm{reflect}\otimes  \ket{v_j}).
\end{equation}
With this we can show that
\begin{equation}
    \widetilde{V}: \ket{0}_\mathrm{reflect}\otimes\ket{j,0} \rightarrow \ket{0}_\mathrm{reflect}\otimes \ket{v_j}.
\end{equation}
The general protocol for implementing $\widetilde{V}$ is clear in \cref{fig:mps_loading}: A controlled state preparation giving $\ket{v_j}$, open-controlled on the phase flipped and Hadamarded ancilla register. 

Conceptually we will do the same thing as above, but with some additional organization. The goal is to implement reflections $(1-2\ket{v}\bra{v})$ defined by the columns $\ket{v} = \ket{(T^{[i]})^{\rchi_i}}$ in \cref{eq:final_t_column}. We achieve this in stages.

\subsection{Reflections over fusion targets}
\begin{figure}[H]
    \centering
    \includegraphics[width=1.0\columnwidth]{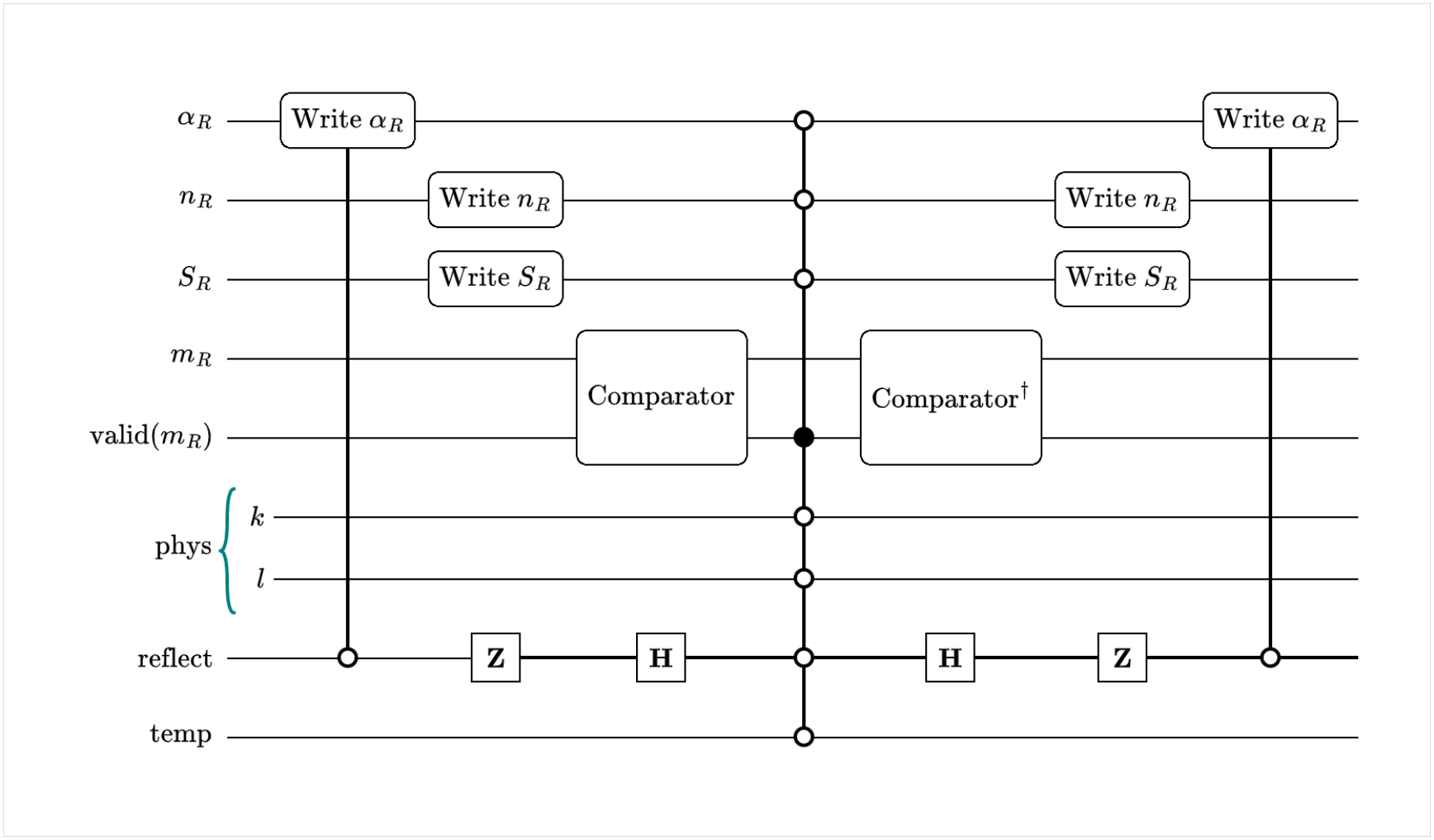}
    \caption{\label{fig:reflect} For every $(\alpha_R,n_R,S_R)$, we do a validity check via a comparator for $-S_R \le m_R \le S_R$, and a reflection about all the registers except the magnetic momenta register, effectively implementing \cref{eq:W_ops}.}    
\end{figure}

At the innermost level, we fix $(\alpha_R,n_R,S_R)$ and implement the reflections over the different $\ket{\alpha_R,n_R,S_R,m_R}$ for $-S_R\le m_R \le S_R$. So we first need a very simple state preparation
\begin{align}
    W_{\alpha_R,n_R,S_R,m_R}^{[i]} : &\ket{0}_\text{reflect} \otimes \ket{0}_\chi \ket{0}_i  \to \nonumber \\
    &\ket{0}_\text{reflect} \otimes \ket{\alpha_R,n_R,S_R,m_R}_\chi \ket{0}_i - \ket{1}_\text{reflect}\otimes \ket{0,n_R,S_R,m_R}_\chi \ket{0}_i =: \ket{E_{\alpha_R,n_R,S_R,m_R}^{[i]}}
\end{align}
Where we have dropped the $\frac{1}{\sqrt{2}}$ normalization factor and will continue doing so going forward. Before the next step, we take the product of the corresponding reflections over $m_R$, obtaining
\begin{align}\label{eq:W_ops}
\prod_{m_R}W_{\alpha_R,n_R,S_R,m_R}^{[i]}(1-2\ket{0}\bra{0})(W_{\alpha_R,n_R,S_R,m_R}^{[i]})^\dagger = 
    \prod_{m_R}(1-2\ket{E^{[i]}_{\alpha_R,n_R,S_R,m_R}}\bra{E^{[i]}_{\alpha_R,n_R,S_R,m_R}})
\end{align}
Of course, the writing of $\alpha_R, n_R,S_R$ needs to only happen for the first and last $W,W^\dagger$ operators, as shown in \cref{fig:reflect}. Furthermore, the reflection can also be done only once, by doing a validity check on $m_R$ satisfying the constraint $-S_R \le m_R \le S_R$. This data allows us to coherently prepare the reduced tensor data, which dominates the cost.

\subsection{Loading a reduced tensor\label{ssec:reducedtensor}}

Next, we need to apply a reflection over
\begin{align}
    \ket{D^{[i]}_{\alpha_R,n_R,S_R,m_R}} := &\ket{0}_\text{reflect} \otimes \ket{\alpha_R,n_R,S_R,m_R}_\chi \ket{0}_i - \nonumber\\ &\ket{1}_\text{reflect}\otimes \sum_{(\alpha_L,n_L,S_L),(n_P,S_P)} (A^{[i]})^{(\alpha_R,n_R,S_R)}_{(\alpha_L,n_L,S_L),(n_P,S_P)}\ket{\alpha_L,n_L,S_L,m_R}_\chi \ket{n_P,S_P,m_{S_P}}_i,
\end{align}
where $m_{S_P}=0$ for $S_P=0$ and $1/2$ otherwise. This is to make sure that the value of the $i$-th register is well-defined, since we need to specify the full three quantum numbers to get a corresponding basis set of two qubits. As a result, the possible values of $\ket{n_P,S_P,m_{S_P}}_i$ are $\ket{00},\ket{10},\ket{11}$, although one could choose a different convention and adapt the circuit later. To create reflections over $\ket{D^{[i]}_{\alpha_R,n_R,S_R,m_R}}$, we need to implement the controlled state preparations:
\begin{align}\label{eq:reducedtensorloading}
    U^{[i]}_{\alpha_R,n_R,S_R} : \hspace{0.5cm} \ket{1}_\mathrm{reflect}\otimes\ket{0, n_R, S_R, m_R}\ket{0} \rightarrow   \hspace{7cm} \nonumber\\ \ket{1}_\mathrm{reflect}\otimes\sum_{(\alpha_L,n_L,S_L),  (n_P,S_P)} (A^{[i]})^{(\alpha_R,n_R,S_R)}_{(\alpha_L,n_L,S_L),(n_P,S_P)}\ket{\alpha_L,n_L,S_L,m_R}_\chi \ket{n_P,S_P,m_{S_P}}_i\hspace{0.3cm} 
\end{align}
This operator conjugates the previous product over $m_R$ and by alternating the conjugation inside, we can show by linear algebra that we are implementing our target reflections
\begin{align}\label{eq:conjugationAtensor}
    &U^{[i]}_{\alpha_R,n_R,S_R} \left(\prod_{m_R}(1-2\ket{E^{[i]}_{\alpha_R,n_R,S_R,m_R}}\bra{E^{[i]}_{\alpha_R,n_R,S_R,m_R}}) \right)(U^{[i]}_{\alpha_R,n_R,S_R})^\dagger = \\
    &(\prod_{m_R}(1-2U^{[i]}_{\alpha_R,n_R,S_R}\ket{E^{[i]}_{\alpha_R,n_R,S_R,m_R}}\bra{E^{[i]}_{\alpha_R,n_R,S_R,m_R}}(U^{[i]}_{\alpha_R,n_R,S_R})^\dagger ) = \\
    &\prod_{m_R}(1-2\ket{D^{[i]}_{\alpha_R,n_R,S_R,m_R}}\bra{D^{[i]}_{\alpha_R,n_R,S_R,m_R}} )
\end{align}

\begin{figure}[H]
    \centering
    \includegraphics[width=1.0\columnwidth]{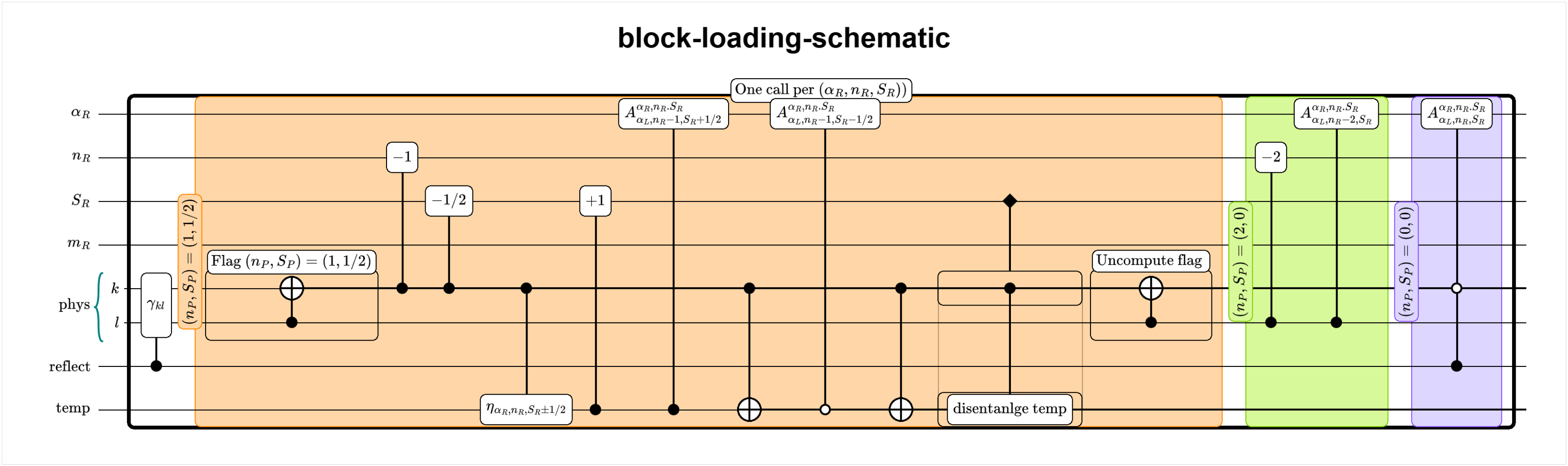}
    \caption{This entire circuit is called once per $(\alpha_R,n_R,S_R)$ to load the corresponding column in \cref{eq:reducedtensorloading}. The notations and subroutines are described in \cref{appsec:reducedtensorloading}. At a high-level, the circuit first creates a superposition over the four different channels depending on $\ket{phys} = \ket{n_P,S_P,m_{S_P}}$, and then for each, loads its corresponding reduced tensor data. A \textit{temp} register will allow to separate the two cases of $S_R\pm 1/2$, later disentangled cheaply thanks to a controlled distinguishing bit (hexagon bit, see more in \cref{appssec:distbit}); the magnetic momenta are corrected in the next stage in \cref{fig:CG}. We note that the subroutines labeled by the reduced tensors $A^{-}_{-}$ are the controlled state preparations, scaling in cost with the number of datapoints.}
    \label{fig:reducedtensorloading}
\end{figure}

In this subroutine we need to both update the value of $\rchi_R=(\alpha_R,n_R,S_R)$ register to allowed $\rchi_L$ values and load the relevant reduced tensor, $A_{\rchi_L=(\alpha_L,n_L,S_L), (n_P, S_P)}^{\rchi_R}$, blocks. A schematic of this state preparation is shown in \cref{fig:reducedtensorloading}, the steps of which are outlined in \cref{appsec:reducedtensorloading}. 

While the above allows for a more targeted and intuitive way of presenting this state preparation, there is an alternative: we can load the entire reduced tensor column data in one shot. The issue is that this state preparation will be over a label $\beta_L$ which ranges over the row size of $B_{n_R,S_R}$. We can force a convention and uncompute the contiguized $\beta_L \to \alpha_L,n_L,S_L$ with a few comparators of logarithmic Toffoli costs and Clifford operations. This implementation, detailed in \cref{appssec:altreducedtensorloading} and \cref{fig:alternativeloading}, while achieving the same asymptotics, practically leads to a lower Toffoli cost and is our basis for later QREs.

\subsection{Applying Clebsch--Gordan Coefficients}
At this stage, all reduced tensors over all $1 \le \alpha_R \le C_{n_R,S_R}$ and $n_R,m_R$ are loaded. Taking the product of previous reflections over all $\alpha_R,n_R$, for a fixed $S_R$, we get
\begin{align}\label{eq:prodD_s}
    \prod_{\alpha_R,n_R,m_R}\left(1-2\ket{D^{[i]}_{\alpha_R,n_R,S_R,m_R}}\bra{D^{[i]}_{\alpha_R,n_R,S_R,m_R}}\right) \ .
\end{align}
The next step is clear: Apply the CG coefficients, and similar to before, we can show by conjugation that we have actually implemented the final target. 

\begin{figure}[h!]
    \centering
    \includegraphics[width=1.0\columnwidth]{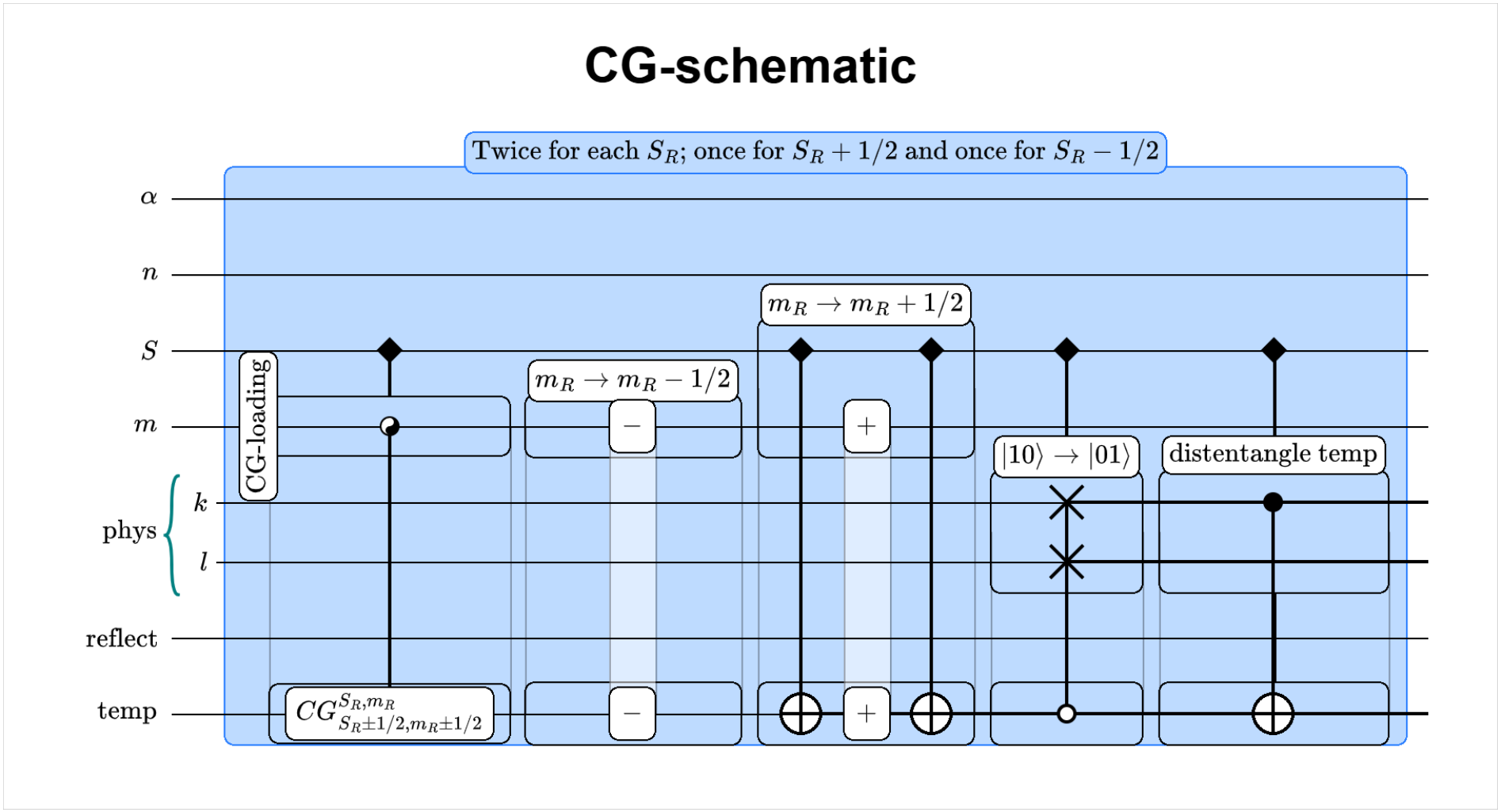}
    \caption{Schematic of CG coefficient loading routine. One call required for each $S_R$ in $\rchi_i$. Where the circuit is repeated for the cases of $S_R\pm1/2$. Since we know the value of $S_R$, we can find a distinguishing bit (\cref{appssec:distbit}) for both cases as denoted by the hexagonal control.
    The yin-yang control
    denotes a multiplexor which selects the relevant value of $m_R$ for each coefficient.}
    \label{fig:CG}
\end{figure}

The Clebsch--Gordan coefficients are dependent on the magnetic sector $m_R, m_L$. The only non-trivial case is when $S_P = 1/2, m_p\in \{\pm1/2 \}$, which is the case when $\ket{phys}\in\{\ket{10}, \ket{01}\}$ in \cref{fig:reducedtensorloading}. Our circuit is shown in \cref{fig:CG}.

We will utilize a temporary qubit to load the two relevant Clebsch--Gordan coefficients which we control on the value of $m_R$ via a QROM. First consider the case $S_L=S_R+1/2$ where we prepare,
\begin{equation}
    \begin{split}
        &\ket{\alpha_L,n_R-1,S_R+1/2, m_R}_{\rchi} \ket{10}_{phys} \ket{0}_{temp} \rightarrow \\
        &CG_{(S_R+1/2, m_R+1/2); (S_P,-1/2)}^{(S_R, m_R)}\ket{\alpha_L, n_R-1, S_R+1/2, m_R}\ket{10}_{phys}\ket{0}_{temp} + \\
        &CG_{(S_R+1/2, m_R-1/2); (S_P,1/2)}^{(S_R, m_R)}\ket{\alpha_L, n_R-1, S_R+1/2, m_R}\ket{10}_{phys} \ket{1}_{temp}
    \end{split}
\end{equation}

We then need to fix-up the registers $\ket{m_R}, \ket{phys}=\ket{kl}$. For the case $m_R\rightarrow m_R-1/2$, we need to subtract 1 in integer form which can be achieved by subtracting the value of the $temp$. For the case $m_R\rightarrow m_R+1/2$, we will add the flipped value of $temp$ controlled on the classically known distinguishing bit of $S_R+1/2$ vs $S_R$ (\cref{appssec:distbit}). Then, controlled on the 0 state of $temp$ we apply SWAP to $phys$ register to correct it for $m_L=m_R+1/2$. Finally we disentangle $temp$ with one Toffoli controlled on the distinguishing bit and $k=1$.  The resulting state is, 
\begin{equation}
    \begin{split}
        CG_{(S_R+1/2, m_R+1/2); (S_P,-1/2)}^{(S_R, m_R)}\ket{\alpha_L, n_R-1, S_R+1/2, m_R+1/2}\ket{01}_{phys} \\ +  CG_{(S_R+1/2, m_R-1/2); (S_P,1/2)}^{(S_R, m_R)}\ket{\alpha_L, n_R-1, S_R+1/2, m_R-1/2}\ket{10}_{phys}.
    \end{split}
\end{equation}

\begin{figure}[H]
    \centering
    \includegraphics[width=1.0\columnwidth]{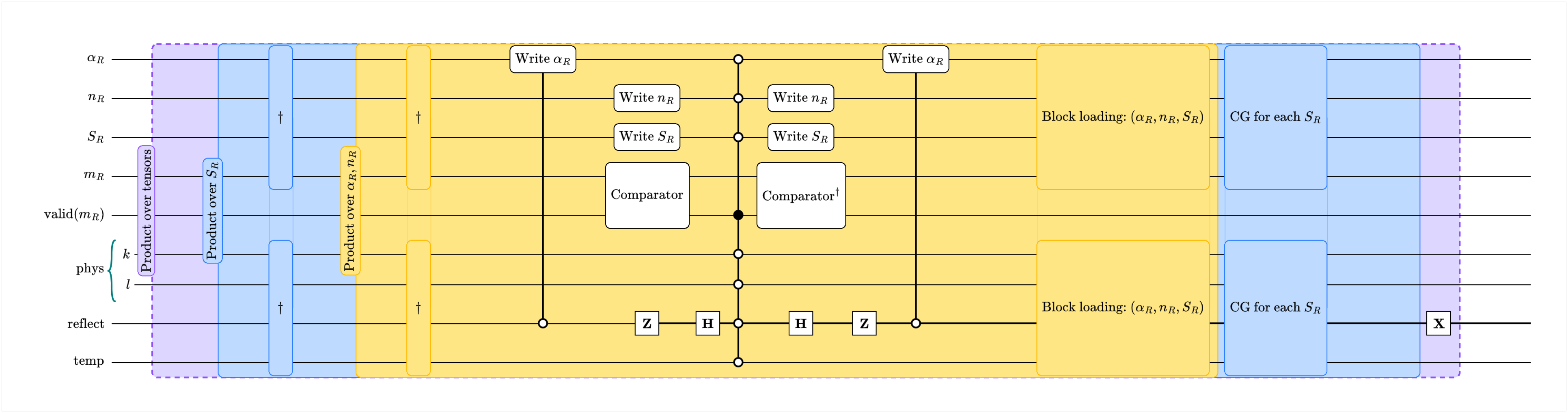}
    \caption{Schematic of the full U(1)~$\times$~SU(2) MPS loading. Proceeding now in order, for each tensor at site $i$, for each quantum spin number occurrence $S_R$, we perform the Clebsh--Gordan subroutine to apply the correct coefficient to each magnetic sector and then load the reduced tensor coefficients, all to implement the full isometry, given by the Householder decomposition, as a sequence of reflections.}
    \label{fig:full_schematic}
\end{figure}

We then repeat this for $S_L=S_R-1/2$. A schematic is presented in \cref{fig:CG}. Note, we only need one call to this subroutine per value of $S_R$ in $\rchi_R$. Thus, we have shown how to implement
\begin{align}
    &CG_{S_R}^{[i]}\ket{D^{[i]}_{\alpha_R,n_R,S_R,m_R}} = \\
    & \ket{0}_a \otimes \ket{\alpha_R,n_R,S_R,m_R}_\chi \ket{0}_i - \nonumber \\
    &\ket{1}_a\otimes \sum_{\substack{(\alpha_L,n_L,S_L,m_L) \nonumber \\(n_P,S_P,m_P)}} CG_{S_Lm_L;S_Pm_P}^{S_Rm_R}(A^{[i]})^{(\alpha_R,n_R,S_R)}_{(\alpha_L,n_L,S_L),(n_P,S_P)}\ket{\alpha_L,n_L,S_L,m_L}_\chi \ket{n_P,S_P,m_{P}}_i
\end{align}
By performing the conjugation, we only need one such application for each different $S_R$ in $\rchi_i$, 
\begin{align}\label{eq:productoveralphaR}
    &CG_{S_R}^{[i]}\left(\prod_{\alpha_R,n_R,m_R}(1-2\ket{D^{[i]}_{\alpha_R,n_R,S_R,m_R}}\bra{D^{[i]}_{\alpha_R,n_R,S_R,m_R}})\right)(CG_{S_R}^{[i]})^\dagger = \\ &\prod_{\alpha_R,n_R,m_R}(1-2\ket{T^{[i]}_{\alpha_R,n_R,S_R,m_R}}\bra{T^{[i]}_{\alpha_R,n_R,S_R,m_R}}) \ ,
\end{align}
giving us the reflections we wanted over \cref{eq:final_t_column}. This last step, the unfolding of the symmetry, is only a fraction ($<1\%$) of the cost in our QREs. A full schematic of the circuit is presented in \cref{fig:full_schematic}.

\section{Resource estimation \label{sec:resourceestimation}}

The main cost of both the baseline dense and this work's SU(2) MPS loading algorithm is LKS state preparation, discussed in \cref{appsec:resestimation}, implemented within the reduced tensor loading subroutine, i.e. \cref{fig:alternativeloading}. 

We first focus our benchmarks on lattice Hamiltonians. Our realistic QREs for Fermi--Hubbard lattice Hamiltonians of small size are instructive in terms of future directions for initial-state preparation on relevant tasks. Similar QREs for molecular systems are also shown, and compared against the recent \cite{rupprecht2026faster}; that paper exploits U(1) symmetry along with a new sparse unitary synthesis protocol that is different from the standard Householder-based one we use.

\subsection{Quantum resource estimate (QRE) settings and quality}
All QREs are for 15 bits of precision for the rotations, consistent with the choice made in \cite{rupprecht2026faster}. The circuits and QREs for both the dense baseline and symmetric loading were implemented and obtained using Workbench \cite{psiquantum_qdk}. This implementation allowed us to do ablation studies, and verify with (sparse) vector simulation that the dense and symmetric loading circuits do reproduce the respective MPS with high fidelity for small Fermi-Hubbard and molecular instances. To get the most accurate QREs, instead of using symbolic expressions, we compiled the \textit{entire} circuit in each instance, and used the Toffoli, qubit and active volume counters of Workbench. This process is inevitably slower than symbolic QREs, but it can be substantially improved by carefully parallelizing the QRE across tensor sites; we verified for even large molecular instances that the counts achieved by parallelized and serial QREs match exactly across multiple QRE metrics, including of course Toffoli, qubit, and active volume. 

\subsection{Benchmarking data}

All MPS tensors were generated through DMRG calculations using \textsc{block2} \cite{zhai2023block2} with SU(2) symmetry enabled. Singlet embedding was utilized for all DMRG calculations with non-zero spin. All MPS tensors were extracted using \textsc{pyblock3} \cite{zhai2021pyblock3}. 

For the comparison between the dense and symmetric loading MPS algorithms, technically, the U(1)~$\times$~SU(2) MPS is densified/unfolded to create the dense MPS, loaded by the dense algorithm, similar to how \cite{rupprecht2026faster} did for unfolding the U(1)-symmetric MPS. Therefore, the dense MPS has a larger bond dimension. For the lattice Hamiltonians, this means a factor of about two to three higher than the SU(2) bond dimension\footnote{As an example, the $4\times 6$ case at 0.75 filling with SU(2) bond dimension 1024 unfolds to a dense with bond dimension 2918.}. For the molecular systems, exact numbers are given later.

\subsection{Fermi--Hubbard}

We first consider the 2D Fermi--Hubbard model on a square-lattice, which is a representative strongly correlated system that is difficult to simulate classically in regimes relevant to quantum simulation. Its Hamiltonian is given by
\begin{equation}
    H=-t\sum_{\langle i,j\rangle,\sigma}
    \left(a_{i\sigma}^{\dagger}a_{j\sigma}
    +a_{j\sigma}^{\dagger}a_{i\sigma}\right)
    +U\sum_i n_{i\uparrow}n_{i\downarrow},
\end{equation}
where $t$ is the nearest-neighbor hopping amplitude, $U$ is the on-site interaction strength,
$n_{i\sigma}=a_{i\sigma}^{\dagger}a_{i\sigma}$, and $\langle i,j\rangle$ indicates summation over nearest-neighbors pairs of lattice sites. We benchmark singlet-sector $4 \times 4$ and $4 \times 6$ lattices with full periodic boundary conditions at $U/t=8$, both at and away from half-filling. As shown in \cref{fig:compress-vs-sweep-fidelity}, direct compression gives lower infidelity than warm-start DMRG at the same target bond dimension for the $4\times 6$, $n=0.75$ case. Details of the MPS generation and compression procedure are provided in \cref{appsec:fermihubbard}.

\subsubsection{Estimating fidelity without the true ground state}

Since direct ground-state overlaps become classically inaccessible at larger sizes, we also test the bond-dimension extrapolation proposed in \cite{berry2025rapid}. Writing $\ket{\rchi}$ for an MPS of bond dimension $\rchi$ and $\ket{\infty}$ for the infinite-bond-dimension limit (i.e., exact diagonalization), the method uses the empirical near-linearity of
\begin{equation}
    \ln\pqty{1-\abs{\braket{\rchi}{\infty}}^2}
    \quad\text{versus}\quad
    \ln\pqty{\rchi}^2
\end{equation}
to estimate the asymptotic fidelity. As seen in \cref{fig:4x44x6-linear-fidelity}, this relationship is approximately linear for the $4\times 4$ and $4\times 6$ Fermi--Hubbard data, with the clearest deviations occurring at very small bond dimensions, roughly $\rchi<10$. Thus, the extrapolation appears useful here provided very small, problem-dependent bond dimensions are excluded from the fit. See more data in \cref{appsec:fidelitiesavtables}.

\subsubsection{QREs}
We plot the results from \cref{appsec:fermihubbardqre} along with some additional relevant datapoints.
\paragraph{The gains from SU(2) MPS loading.} For both the dense and SU(2) case, we estimated the Toffoli, active volume, and qubit count of each subroutine of our circuits.
The SU(2) circuit used for reduced tensor loading is the previously mentioned alternative in \cref{ssec:reducedtensor} with details in \cref{appssec:altreducedtensorloading}.

\begin{figure}[h!]
    \centering
    \includegraphics[width=0.675\linewidth]{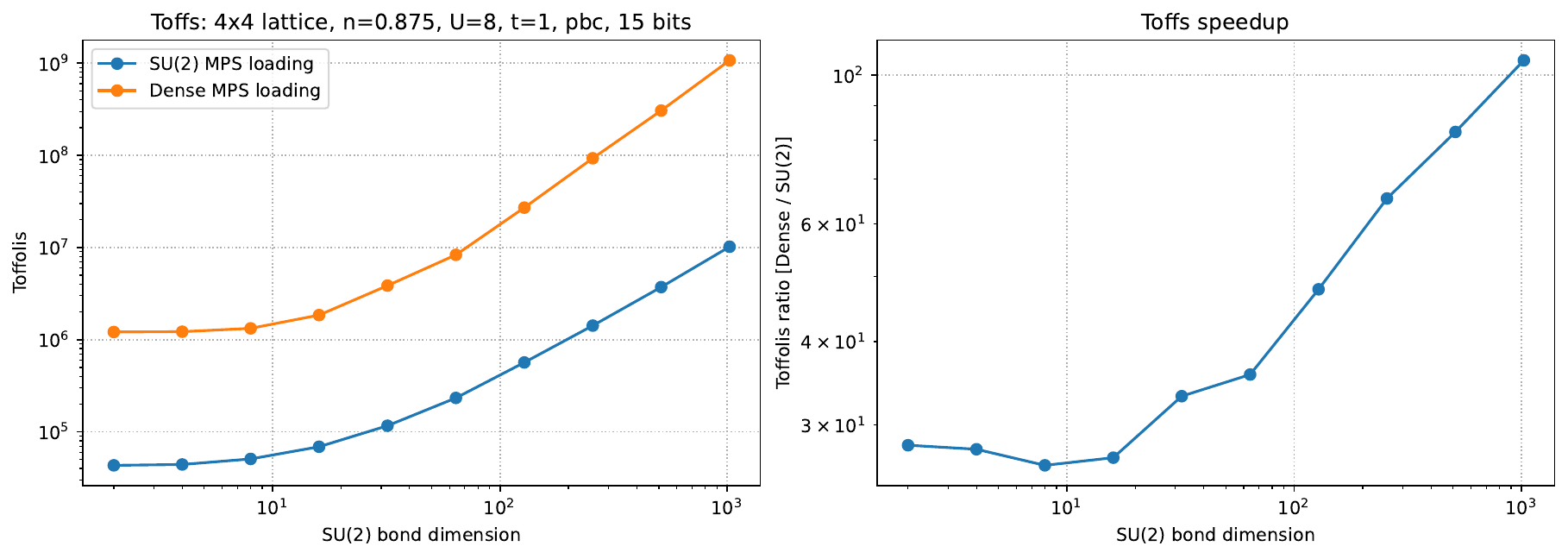}
    \includegraphics[width=0.3\linewidth]{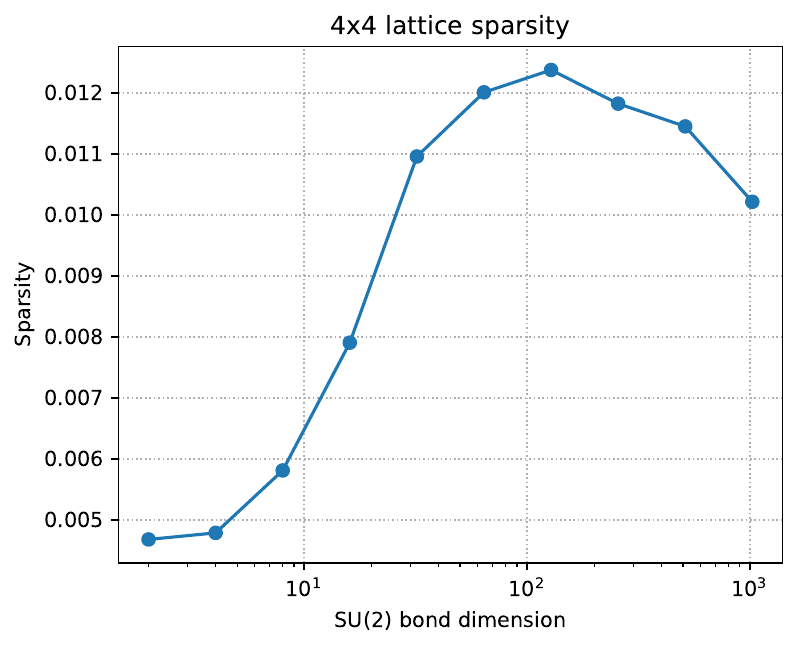}
    \includegraphics[width=0.75\linewidth]{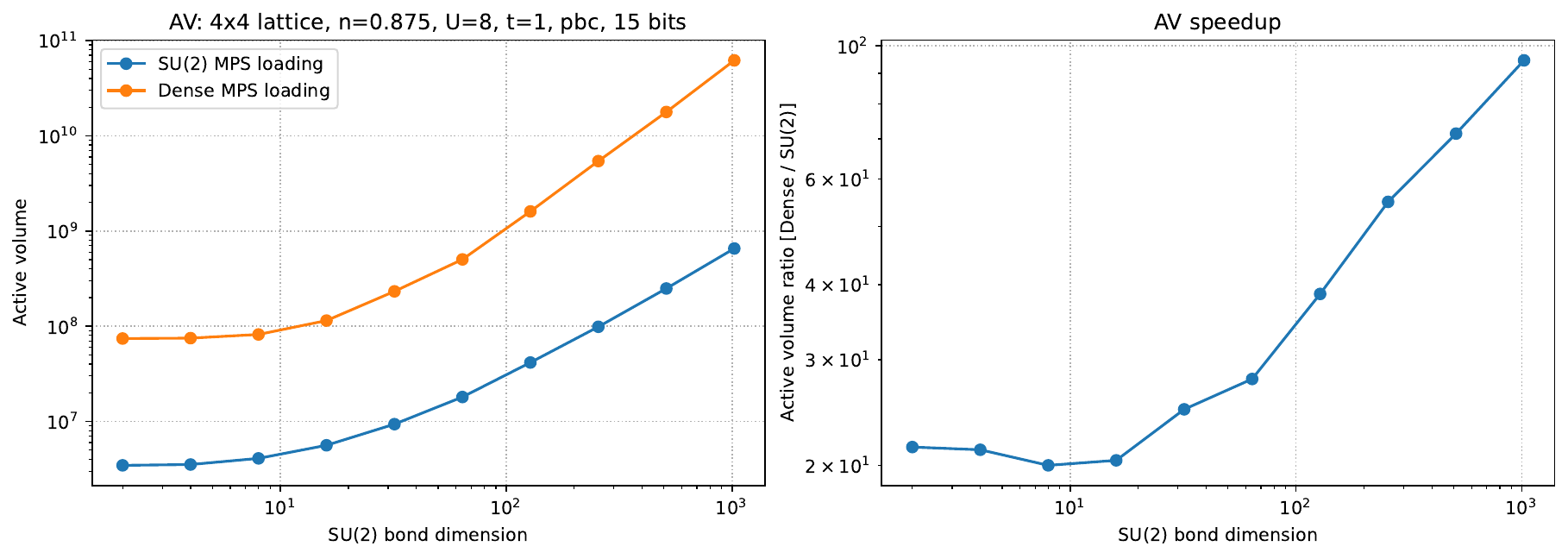}
    \vspace{0.5em}
    \noindent\rule{\linewidth}{0.4pt}
    \vspace{0.5em}
    \includegraphics[width=0.675\linewidth]{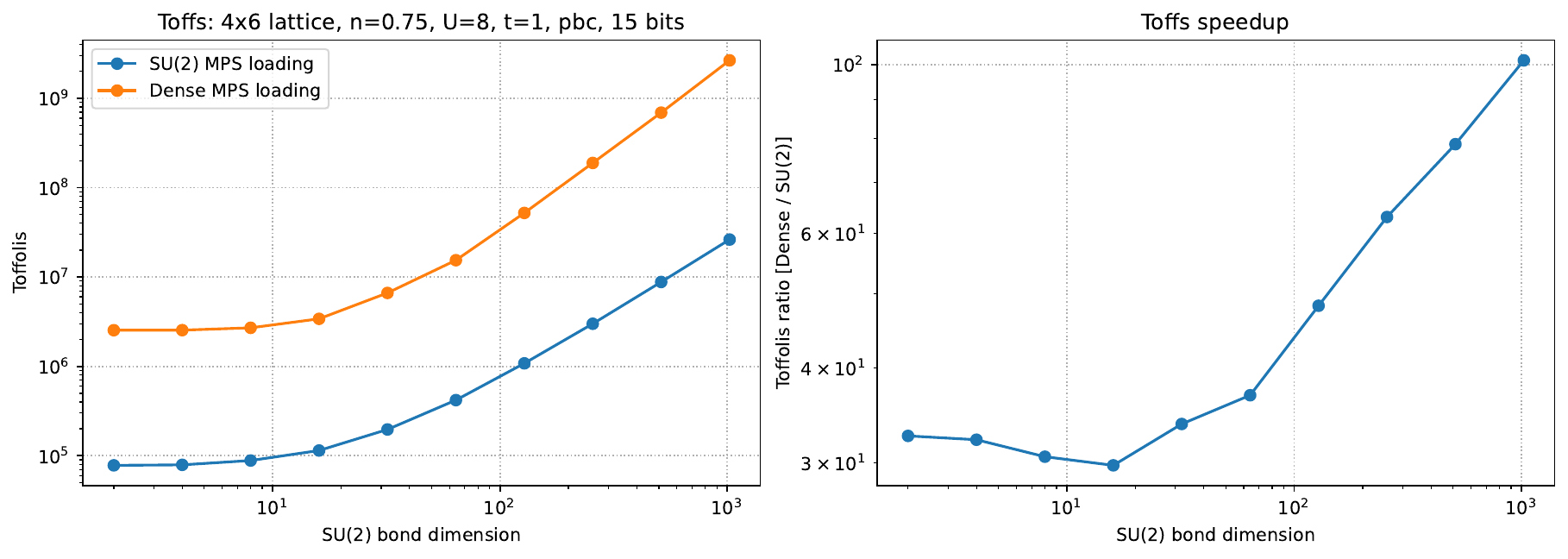}
    \includegraphics[width=0.3\linewidth]{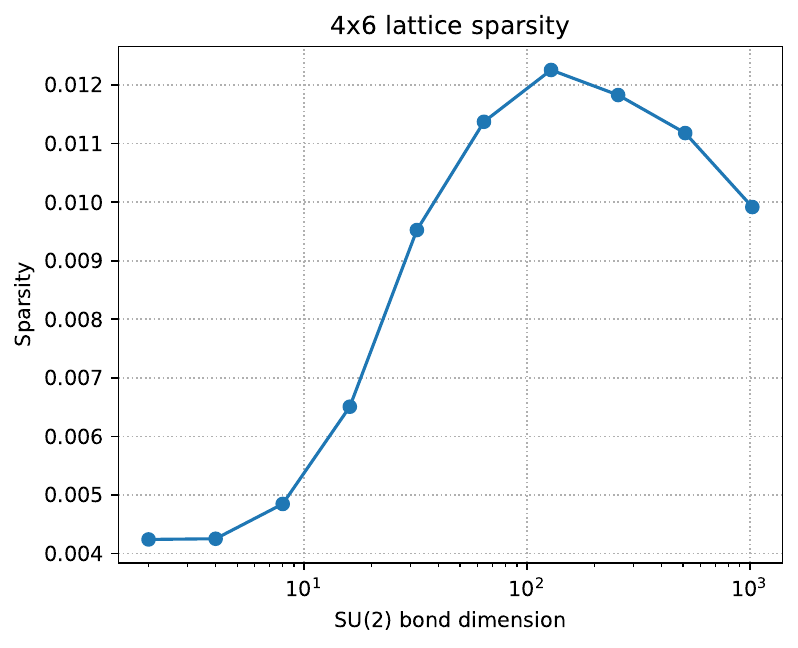}
    \includegraphics[width=0.75\linewidth]{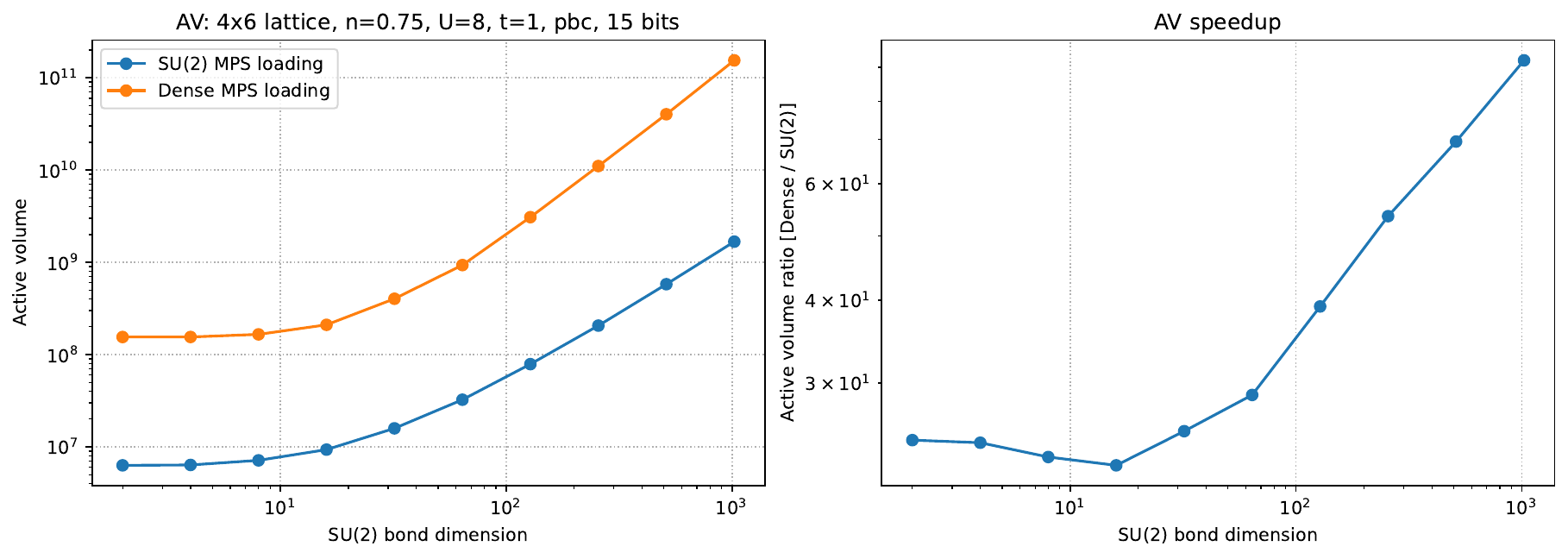}
    
    \caption{QREs for loading a SU(2) MPS for the Fermi--Hubbard model. Plots on top are for the $4\times 4$ at $1/8$-doping, plotting the Toffoli costs, speedup ratio, and sparsity. Similarly for the bottom plots for $4\times 6$ at filling $0.75$. Note the dense loading is a loading of the densified symmetric MPS, thus its bond dimension is larger than the corresponding symmetric MPS. The loading costs, plotted here without any QROM qubit-gate tradeoff, asymptotically tracks the number of datapoints up to the same constant factor as they both use the same Householder-LKS-based method. Thus, as the size of the MPS gets larger, the LKS costs become dominant (see \cref{appsec:fermihubbardqre}), and the speedup ratio approaches the inverse sparsity, which is about $100=(1\%)^{-1}$ in examples shown above. Sparsity is measured as the total number of nonzero ($\abs{x}>10^{-15}$) elements in the SU(2) MPS over the sum of the size of the dense MPS tensors (see more details on how this should be compared to the speedup ratio in \cref{appsec:densitydefinition}). This tracks the sparsity results shown in \cite{levy2020distributed}.}
    \label{fig:fh_toff_sparsity}
\end{figure}

\begin{figure}[h!]
    \centering
    \includegraphics[width=0.75\linewidth]{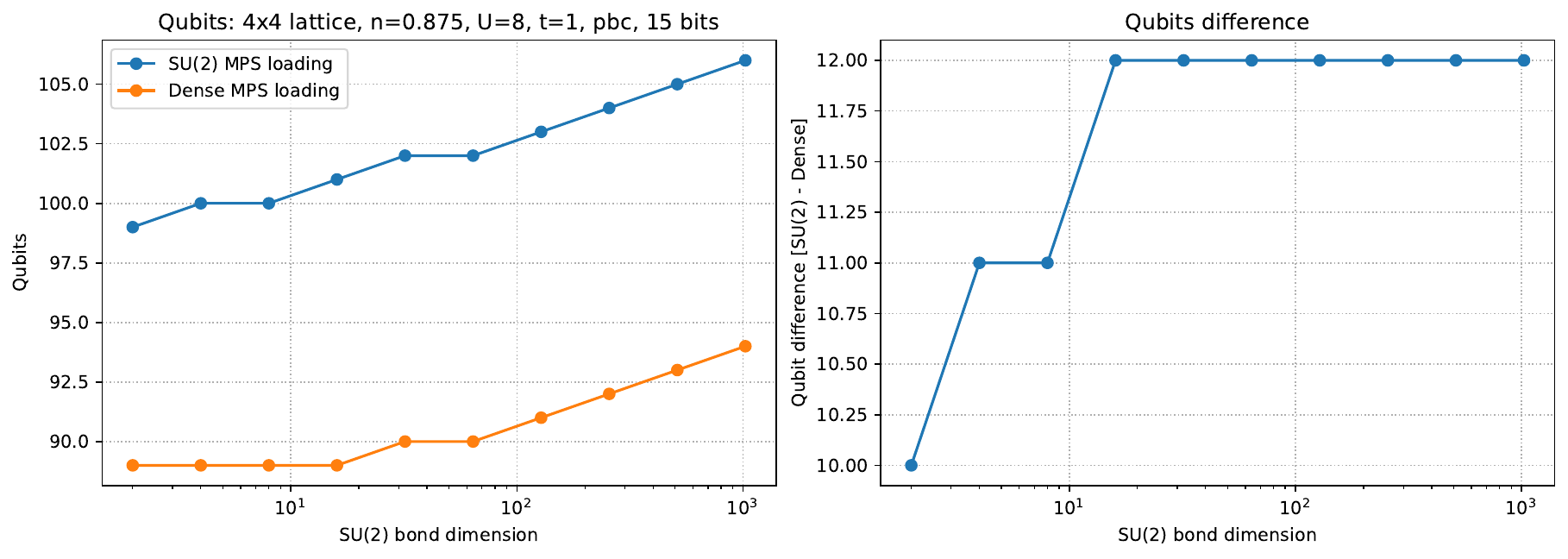}
    \includegraphics[width=0.75\linewidth]{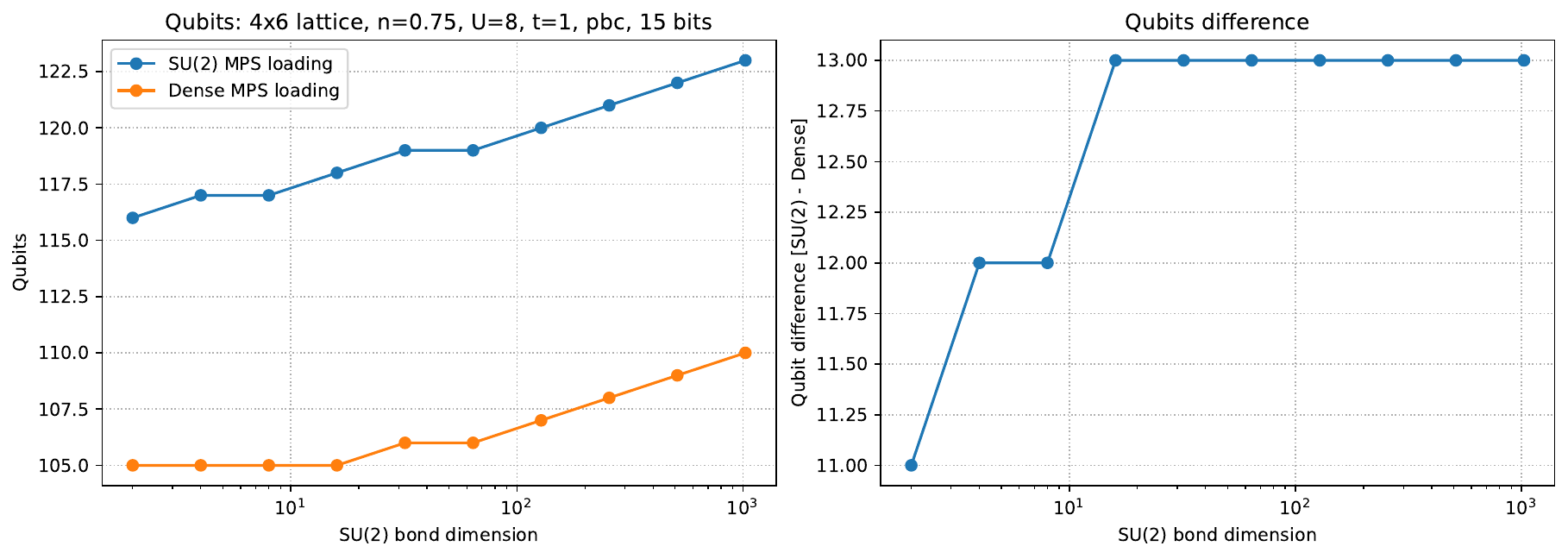}
    \caption{Qubit counts between the loading algorithms are not so different, showcasing SU(2) MPS loading improvement comes at essentially negligible qubit expense. Plots are shown for the same instances as in \cref{fig:fh_toff_sparsity}. Although not shown in these examples (but see \cref{tab:chemical_mps_loading_qre}), note that the qubit count picture generally changes to the \textit{benefit} of the SU(2) MPS loading when selecting the QROM-gate-optimal tradeoff, as the dense loading MPS deals with larger state preparations, thus consuming more qubits to achieve optimal gate count; on the other hand, this also means dense MPS benefits more in gate reduction than SU(2) MPS (see relevant numerics in \cref{ssec:molecular_sys}).}
    \label{fig:fh_qubits}
\end{figure}

The trend clearly shows how the sparsity, measured by $s_{\mathrm{SU(2)}}$ as the total number of nonzero ($\abs{x}>10^{-15}$) elements in the symmetric MPS over the sum of the size of the dense MPS tensors, significantly lowers the cost (\cref{fig:fh_toff_sparsity}) by multiples of $10\times$, with an overhead that is largely dominated by the LKS costs in both cases (\cref{appsec:fermihubbardqre}). The qubit counts are also competitive (\cref{fig:fh_qubits}). 

\paragraph{The worrying trend of cost.}
Despite the success here in reducing the cost compared to the baseline, we need to note that the costs are still quite high. The lattice size required for quantum advantage of the Fermi--Hubbard model is usually large. For example, for ground-state energy estimation, notable results include PEPS \cite{liu2025accuratesimulationhubbardmodel} with lattice sizes up to $16 \times 16$, which means a larger $20 \times 20$ may be the next frontier \cite{bartschi2024potential}. A $20\times 20$ lattice would require 800 qubits just to encode the system register. In general, without significant innovation on the encoding rate of the system register, it is likely that early-generation fault-tolerant machines may have to use the vast majority of their qubit capacity to simply encode the system, limiting any room for gate-qubit tradeoff, as opposed to molecular instances. On the other hand, as the size of the system grows, so does the required bond dimension to ensure high-enough fidelity with the desired eigenstate (see \cref{sec:futuredirections,appsec:fidelitiesavtables} and \cite{liu2025accuratesimulationhubbardmodel}). For larger lattice systems than shown here (beyond $4\times 6$), one may well have to consider loading a DMRG with a few thousands to tens-of-thousands bond dimension. This raises the question of whether the loading costs are justified, especially when it has been shown that QPE for ground-state energy estimation on these systems is on the order of $10^5-10^6$ Toffolis \cite{kan2024resource,Apel2026FH}.

In fact, looking deeper into the DMRG data, we see that the reduced tensors are actually dense themselves, and a truncation of $10^{-8}$ on nonzero elements fails to establish any new significant sparsity. More precisely, for a $4 \times 6$ system at 0.75 filling with bond dimension 1,000, the sum of the sizes of the reduced tensors (i.e. $\sum_i \dim_{\mathrm{SU(2)}} A^{[i]}$) is 4,056,905, which is also equal to the number of nonzero elements with absolute value $>10^{-15}$ in the reduced tensors $A^{[i]}$, and 4,056,881 many with absolute value $> 10^{-8}$. Going to a system of size $20\times 20$ would require a much larger bond dimension for sufficient fidelity. With a highly optimistic bond dimension of 4,000, multiplying by $\sim 16$ the number of nonzero elements, and given the increased number of sites, proportionally increasing it further by another $\sim 16.6$, totals almost a billion data points. Reiterating the point of no room for qubit-gate tradeoff, and assuming a highly efficient unitary synthesis protocol matching the Toffoli count with the number of data points without qubit tradeoff\footnote{The new protocols in \cite{berry2025rapid, rupprecht2026faster} do not achieve such a general result for our reduced tensors isometries which have arbitrary dimensions. So this is a hypothetical.}, the result is $\sim 10^9$ Toffolis. Assuming the early-generation machine can accommodate such gate count, this is a situation that one would still likely want to avoid, where the loading of a classical data is somehow 1,000 times more difficult than the core part of the quantum algorithm (QPE) that is supposed to showcase the quantum advantage. This does not take into account the classical cost of preparing such high quality MPS, which at large lattice size and bond dimension, can be rather slow, costly, and difficult to obtain by itself.

In light of these, while we do discuss how to improve the MPS loading costs further, we also reconsider whether DMRG-based MPS is a feasible initial-state preparation modality for lattice Hamiltonian tasks, and discuss alternatives in \cref{sec:futuredirections}, noting that ultimately, no modality is likely to be a winner in all scenarios.

\subsection{Molecular systems}\label{ssec:molecular_sys}

We also evaluate the SU(2) MPS loading algorithm for several prototypical strongly correlated and biologically relevant systems: the heme group in the active site of cytochrome P450 and two iron-sulfur clusters, \ce{[Fe2S2]^2-} and \ce{[Fe4S4]^2-}. Cytochrome P450 is an enzyme that plays a crucial role in drug metabolism, whereas the iron-sulfur clusters serve as models for the FeMoco cofactor that participates in nitrogen fixation. The P450 heme cofactor and iron-sulfur clusters are commonly used as benchmark systems for quantum computing algorithms, including ground-state energy estimation \cite{low2025fast, goings2022reliably, caesura2025faster, rocca2024reducing} and initial-state preparation \cite{ollitrault2024enhancing, berry2025rapid, lee2023evaluating, fomichev2024initial}.

For P450, we used the same active-space Hamiltonian as in \cite{caesura2025faster, goings2022reliably, rupprecht2026faster}, corresponding to a (63e, 58o) active space in the cc-pVDZ basis set \cite{balabanov2005systematically, balabanov2006basis, woon1993gaussian}. We computed the ground-state wave function for the high-spin ($S=5/2$) and low-spin ($S=1/2$) states with a SU(2) bond dimension of 1,500. We used the same active-space Hamiltonians as in \cite{rupprecht2026faster} for both \ce{[Fe2S2]^2-} and \ce{[Fe4S4]^2-}, which are of size (30e, 20o) and (54e, 36o) respectively. We computed the ground-state singlet with a SU(2) bond dimension of 1,000 for both systems. Though, we do not estimate the fidelity for these systems since they are more complicated (see \cite{berry2025rapid}).

\paragraph{A note on the P450 inputs for QRE.}
The P450 calculations used singlet embedding, which introduces an
auxiliary spin to represent a physical non-singlet state within a singlet calculation. We first removed this auxiliary embedding to recover the physical 58-orbital states with $S=1/2$ and $S=5/2$. The SU(2) bond dimension of these extracted states is then no longer $1500$: their largest reduced bonds are 2816 and 6490, respectively. To impose a common total reduced bond cap of 1500, we applied a symmetry-preserving singular-value truncation across the spin sectors at each bond, then normalized the states and restored canonical form. The dense and SU(2) resource estimates use these same compressed physical states. Unfolding them into dense tensors introduces no further bond truncation. The squared overlaps with the corresponding normalized, uncompressed supplied states are $>99.9\%$ for the doublet and sextet. These fidelities quantify the additional compression error relative to the original DMRG states and not to the exact ground states, which we do not have access to.

\subsubsection{QREs}
We comparing the dense and SU(2) MPS loading in \cref{tab:chemical_mps_loading_qre}.

The gate and even volume cost reduction is reduced when allowing for qubit-gate tradeoff compared to the lattice Hamiltonian examples at similar bond dimensions. Indeed, what gives SU(2) MPS loading its advantage are state preparations of typical size , say, $n_{\mathrm{SU(2)}}$ that are smaller than the typical $n_{\mathrm{dense}}$ in the dense case. However, this also reduces the gate cost by a smaller factor at the optimal trade-off point, since $\sqrt{n_{\mathrm{SU(2)}}}<\sqrt{n_{\mathrm{dense}}}$, though using less qubits as well to reach the optimal gate count. Either way, if using a volume measure, the reductions even in the gate-optimal case are still significant, reaching $\sim 140\times$ volume reduction in the case of \ce{[Fe4S4]^2-}. 

The QREs for our systems are compared with \cite{rupprecht2026faster}. While we note that our DMRGs did not necessarily reproduce the same exact data, they did achieve very close energies and our densified symmetric MPS has a bond dimension that is slightly different: 3,302, 7,246, 4,116, 7,974 compared to their 3,302, 6,832, 4,751, 8,278.

\begin{table}[H]
\centering
\resizebox{\textwidth}{!}{%
\begin{tabular}{llrrrrrrrrrrrrrrrr}
\toprule
System & $(N_e,N_o)$ & $M$ & $s_{\rm SU(2)}$ & $s_{\rm SU(2)}^{-1}$ & \multicolumn{3}{c}{SU(2), $\lambda=1$} & \multicolumn{3}{c}{Toffolis, QROM-opt.} & \multicolumn{3}{c}{Active volume, QROM-opt.} & \multicolumn{2}{c}{Qubits, QROM-opt.} & \multicolumn{2}{c}{\shortstack{SU(2) reduced Toffolis /\\total Toffolis}} \\
\cmidrule(lr){6-8}\cmidrule(lr){9-11}\cmidrule(lr){12-14}\cmidrule(lr){15-16}\cmidrule(lr){17-18}
& & & & & Toffolis & Active volume & Qubits & dense & SU(2) & speedup & dense & SU(2) & speedup & dense & SU(2) & $\lambda=1$ & QROM-opt. \\
\midrule
{}[Fe$_2$S$_2$]$^{-2}$ & $(30,20)$ & $1000$ & $8.84\times10^{-3}$ & $113.1$ & $1.96\times10^{7}$ & $1.23\times10^{9}$ & $113$ & $2.10\times10^{8}$ & $1.04\times10^{7}$ & $20.28\times$ & $1.92\times10^{10}$ & $7.39\times10^{8}$ & $25.98\times$ & $596$ & $223$ & $98.45\%$ & $97.08\%$ \\
{}[Fe$_4$S$_4$]$^{-2}$ & $(54,36)$ & $1000$ & $2.26\times10^{-3}$ & $442.2$ & $5.50\times10^{7}$ & $3.44\times10^{9}$ & $149$ & $1.55\times10^{9}$ & $2.83\times10^{7}$ & $54.82\times$ & $1.76\times10^{11}$ & $2.02\times10^{9}$ & $87.05\times$ & $662$ & $259$ & $98.38\%$ & $96.86\%$ \\
P450 ($S=1/2$) & $(63,58)$ & $1500$ & $1.14\times10^{-2}$ & $87.4$ & $2.15\times10^{8}$ & $1.33\times10^{10}$ & $193$ & $1.45\times10^{9}$ & $9.98\times10^{7}$ & $14.56\times$ & $1.48\times10^{11}$ & $7.14\times10^{9}$ & $20.66\times$ & $706$ & $303$ & $98.79\%$ & $97.39\%$ \\
P450 ($S=5/2$) & $(63,58)$ & $1500$ & $3.81\times10^{-3}$ & $262.6$ & $2.04\times10^{8}$ & $1.26\times10^{10}$ & $192$ & $3.10\times10^{9}$ & $9.63\times10^{7}$ & $32.15\times$ & $3.56\times10^{11}$ & $6.84\times10^{9}$ & $52.04\times$ & $706$ & $302$ & $98.73\%$ & $97.32\%$ \\
\bottomrule
\end{tabular}}
\caption{Our MPS-loading QREs with 15 bits of precision for the rotations, for systems with $N_e$ electrons and $N_o$ spatial orbitals, and SU(2) bond dimension $M$. We define $s_{\text{SU}(2)}$ as in \cref{fig:fh_toff_sparsity}, with a corresponding 1e-15 threshold. Higher thresholds did not materially increase the sparsity $s_{\text{SU}(2)}^{-1}$. The reduced-loading fractions represent how the total cost is influenced by the number of datapoints through the LKS state preparations. We show the Toffoli/active volume costs with (QROM-opt.) and without ($\lambda=1$) qubit tradeoff for the symmetric loading. Speedup is expected to be reduced in the QROM-optimal setting, as dense loading can leverage its main disadvantage of larger state preparation to its benefit, by utilizing more qubits to get higher gate reduction factor.}
\label{tab:chemical_mps_loading_qre}
\end{table}

\begin{table}[H]
\centering
\resizebox{\textwidth}{!}{%
\begin{tabular}{lrrrrrrrrrrrrrr}
\toprule
System & our $\rho^{-1}$ & prior $\rho^{-1}$ & \multicolumn{3}{c}{Our QROM-opt. Toffolis} & \multicolumn{2}{c}{Our QROM-opt. qubits} & \multicolumn{3}{c}{Prior Toffolis} & \multicolumn{2}{c}{Prior qubits} & \multicolumn{2}{c}{Volume reduction} \\
\cmidrule(lr){4-6}\cmidrule(lr){7-8}\cmidrule(lr){9-11}\cmidrule(lr){12-13}\cmidrule(lr){14-15}
& & & dense & SU(2) & speedup & dense & SU(2) & dense & sparse & speedup & dense & sparse & ours & prior \\
\midrule
{}[Fe$_2$S$_2$]$^{-2}$ & $29.1$ & $23.3$ & $2.10\times10^{8}$ & $1.04\times10^{7}$ & $20.28\times$ & $596$ & $223$ & $8.3\times10^{7}$ & $5.4\times10^{6}$ & $15.4\times$ & $545$ & $544$ & $54.19\times$ & $15.4\times$ \\
{}[Fe$_4$S$_4$]$^{-2}$ & $51.8$ & $42.5$ & $1.55\times10^{9}$ & $2.83\times10^{7}$ & $54.82\times$ & $662$ & $259$ & $5.3\times10^{8}$ & $2\times10^{7}$ & $26.5\times$ & $1052$ & $576$ & $140.11\times$ & $48.4\times$ \\
P450 ($S=1/2$) & $26.9$ & $22.5$ & $1.45\times10^{9}$ & $9.98\times10^{7}$ & $14.56\times$ & $706$ & $303$ & $6.2\times10^{8}$ & $3.8\times10^{7}$ & $16.3\times$ & $1034$ & $620$ & $33.94\times$ & $27.2\times$ \\
P450 ($S=5/2$) & $47.0$ & $32.8$ & $3.10\times10^{9}$ & $9.63\times10^{7}$ & $32.15\times$ & $706$ & $302$ & $1.1\times10^{9}$ & $4.7\times10^{7}$ & $23.4\times$ & $1074$ & $620$ & $75.15\times$ & $40.5\times$ \\
\bottomrule
\end{tabular}}
\caption{Comparison with the inverse-density, \emph{dense}, and \emph{sparse} QRE entries of \cite[Tab. 1]{rupprecht2026faster}. While gate reductions are similar, qubit reductions are more significant in our symmetric MPS loading. We make a clarification on the definition of density and how that should be compared to the speedup ratios in \cref{appsec:densitydefinition}.}
\label{tab:chemical_mps_loading_prior}
\end{table}

These slight differences should not cause major differences in QREs for the baseline dense algorithms that are the same, but comparing our QROM-opt. and dense numbers in \cref{tab:chemical_mps_loading_qre} and those of the prior work in \cref{tab:chemical_mps_loading_prior}, we see that our QREs are higher than the literature on gates, while lower on qubits.

Our new loading algorithm that leverages the SU(2) structure uses about twice as many gates compared to \cite{rupprecht2026faster} while operating on about half of the qubits. Thus volume-wise we see a competitive result, but we also note that we are using an older unitary synthesis method. As sketched in \cref{sec:futuredirections}, more efficient unitary synthesis such as \cite{berry2025rapid} can bring down the gate cost by multiples, while likely preserving the qubit count as their cost equations present qubit-gate tradeoff ratios similar to LKS. Either way, despite the use of an older synthesis method, our gate counts are not so different, most likely due to the greater sparsity provided by the additional SU(2) symmetry. 

In terms of the reduction brought from the dense case, as shown in \cref{tab:chemical_mps_loading_prior}. For the example of \ce{[Fe4S4]^2-}, the sparse synthesis method applied on U(1) tensors in \cite{rupprecht2026faster} shows $48.4\times $ volume reduction, compared to our $140 \times$. In general, we observe similar gate reduction for the QROM-optimal instances compare to prior work, but higher volume reduction in all instances.

Similar to the previous section on lattice Hamiltonians, we ask whether DMRG-based SU(2) MPS is a feasible initial state modality for molecular systems. While we showed that for lattice Hamiltonians the cost of MPS loading may be substantially larger than any interesting downstream task; here, we note that the largest Toffoli counts for the well-known benchmark of P450 stands at about a tenth of a downstream QPE cost (see \cite{low2025fast} for more). But this is not necessarily the pattern, as in fact the other two instances have a slightly lower QPE cost and one would have to see how future improvements on both front increase or decrease this gap.

It is also instructive to look at the number of nonzero parameters in SU(2) for a Toffoli lower bound, assuming an attempt to load the MPS without any tradeoff. This could be perhaps for the more ambitious goal of running a QPE on a machine with a qubit count that is close to the system register size. Note that this in turn would likely increase the QPE cost as well. The number of nonzero parameters for the SU(2) MPS' above are 2,849,975, 8,221,013, 32,736,934, and 29,146,873 corresponding to \ce{[Fe2S2]^2-}, \ce{[Fe4S4]^2-}, and the P450s at spin $1/2$ and $5/2$, respectively. As a reference, the first two have QPE cost $4\times 10^7$ and $1.7\times 10^7$ and the last two about $4\times 10^8$ \cite{low2025fast}. Once again, we note that these QPE costs are using high qubit counts of 463, 868, and 1,150 for the P450 systems. For \ce{[Fe4S4]^2-} (and P450), they are on the order of (a few) $10^7$ data points in the MPS, pointing to an as-large a Toffoli count assuming no trade-off. We can only conclude that to achieve said ambitious goal, one should actively look for other ways of compressing the MPS, or alternatives to (DMRG-based) MPS altogether.

\section{Direct unitary synthesis}

Alternatives to the Householder-LKS method, such as \cite{berry2025rapid,rupprecht2026faster}, can be used for SU(2) MPS. In the Householder-based method the reflections restrict the action to the required Hilbert space. A directly synthesized unitary, by contrast, acts on the whole register, so the subspaces on which it must act correctly have to be specified explicitly. We can keep the incoming sector labels $(n_R,S_R)$ untouched as multiplexer controls and write the fusion channel into the physical register in a \emph{coupled} code that distinguishes $S_L=S_R-1/2$ from $S_L=S_R+1/2$. In this basis the reduced site isometry is exactly block diagonal over the sectors $(n_R,S_R)$, so the synthesis methods apply to every block, with the synthesis layers of all blocks merged. A sector-independent fusion update of the $(n,S)$ registers and a single data-independent Clebsch--Gordan rotation in the singly-occupied physical subspace, multiplexed on $(S_L,m_R)$, then complete the site isometry. In terms of speedup, while the circuit generally simplifies, e.g. neither a reflection ancilla nor a temporary qubit is needed, it is not quite clear how much one can expect.

Note that while there are four fusion channels, their multiplicities need not equal the incoming multiplicity: the aspect ratio $M_s/C_s$ can exceed four, and the padding by zero obviously only increases this ratio before becoming an integer. This becomes important since e.g. \cite{berry2025rapid} claims a cost reduction of about 3.5 for $d=4$ for the typical $\rchi d \times \rchi$ dimensional isometry in the dense case, and if $d$ grows, then the gains compared to the Householder-LKS method stop at around $d=30$. Nevertheless, some speedup in between is expected. More details on the scaffolding that makes these direct unitary syntheses method a simple plug-in is provided in \cref{appsec:direct}.

\section{Future directions}\label{sec:futuredirections}

Given that our scheme above was for a product of an abelian and non-abelian symmetry group, we consider generalization of the circuit to any symmetry group of the form $G_1\times \cdots \times G_k$ to be feasible with the same scheme, building up the quantum numbers one by one and performing the state preparation for the reduced tensor blocks.

While this is an interesting abstract exercise, it could also be of practical significance as larger symmetries lead to fewer data points. But at least for molecular systems, we may not have additional symmetries that apply to nearly as wide a selection of systems as charge-spin symmetry does; except for perhaps point-group symmetries. Further, before considering implementing the quantum algorithm for any symmetry, we have to also consider the support and availability of corresponding DMRG software tools.

One may explore alternatives to MPS, particularly other tensor-network states. PEPS is likely the first option, but their loading onto a quantum computer requires substantial research. First, PEPS are among the class of tensor networks that cannot be exactly canonicalized in general, unlike an MPS. But there are variational techniques that canonicalize PEPS, while potentially maintaining the quality \cite{haghshenas2019conversion}. 

On the other hand, PEPS are notoriously hard to optimize and require supervised optimization by an expert. Indeed, as previously mentioned, these computational tasks currently take significant time and classical computing resources: \cite{liu2025accuratesimulationhubbardmodel}'s PEPS optimizations runtime is on the order of weeks, making this the bottleneck of the computational pipeline. In conclusion, PEPS, even if used efficiently, may be hard to incorporate in a streamlined, automated workflow.

Another alternative to MPS is the sum of Slater determinant state (SOS) \cite{fomichev2024initial}, which is relatively easy to load into quantum computer as an initial state. Numerical algorithms with SOS as the ground state are usually a rougher approximation than DMRG, but scale better with system size. These include self-consistent Hartree-Fock (HF) methods, prominently coupled cluster singles and doubles (CCSD), semistochastic heat-bath configuration integration (SHCI), etc. 
Given the low Toffoli cost in QPE for large lattice Hamiltonians (\cite{kan2024resource}), it is reasonable to first attempt the simplest Slater-determinant-based initial states method. Hartree-Fock provides the simplest example: it produces a single Slater determinant, can be optimized classically on substantially larger lattices than high-bond-dimension tensor-network states, and has a method to convert directly into tensor-network states. Systematic extensions include symmetry-projected mean-field states \cite{khinevich2026symmetry}, small linear combinations of symmetry-related determinants, and selected-configuration expansions. These approaches have a systematically improved correlation and symmetry restoration with a controlled increase in state-preparation cost. They may also be combined with low-bond-dimension MPS, for example by using a compact multi-determinant expansion to capture correlations that are otherwise costly to represent within the MPS alone. While many of these avenues may already have been somewhat explored in the literature, more numerics and optimized quantum resources are required for comparison.

Taking a step back and looking at the general initial state problem, we have to recall that one can never prove the fidelity of the initial state with the ground state of a general molecular or lattice Hamiltonian. And while one could invest efforts on a select set of systems and perhaps prove some kind of lower bound on the overlap, we believe the simple fact that a scalable workflow should compare multiple classes of classical initial states and also quantum state-preparation strategies, rather than assuming that the most accurate classically accessible state is necessarily the optimal state to prepare on the quantum computer.

\section*{Acknowledgments}
All authors contributed to the writing and editing of the manuscript. This work is more than one year old, and since then, the circuits and the main conclusions were shared with many of our colleagues in useful discussions. In particular, we would like to credit Sam Pallister for the use of the comparator for the reflection over fusion targets, Harriet Apel and Jessica Lemieux for a review of the oldest draft and useful discussions, as well as Mark Steudtner for a review of a more recent draft. Construct's Circuit Designer was used to draw some of the quantum circuit diagrams\footnote{freely available at \url{https://construct.psiquantum.com/cd/}.}.

\section*{Declaration of AI usage}
The authors take full responsibility for the content of this paper. Artificial Intelligence was used in making some of the graphics, debugging the code and circuits on some missing controls, as well as to fill in the gaps of the sketch we had for the loading with direct unitary synthesis, written in \cref{appsec:direct}. Besides these, the authors made the rest of the content including the data generation, fidelity and all other relevant calculations for preprocessing for the quantum circuit, and the ideation, synthesis, verification, and Workbench implementation of the circuits. 

\bibliographystyle{alpha}
\bibliography{main}

\begin{appendices}

\section{Reduced tensor loading steps}\label[appsec]{appsec:reducedtensorloading}

\begin{enumerate}
    \item We first create a superposition over the relevant states of the physical register such that on the $\ket{1}$ branch of the reflection ancilla,
    \begin{align}
        \ket{1}_{\text{reflect}}\otimes \ket{0,n_R,S_R,m_R}_{\rchi}(\gamma_{00} \ket{00}_i +\gamma_{10} \ket{10}_i +\gamma_{11} \ket{11}_i),
    \end{align}
    where the values of $\gamma_{n_P,S_P,m_{S_P}}$ are given by $\sqrt{\sum_{(\alpha_L,n_L,S_L) \in \rchi_{i-1}}\abs{(A^{[i]})^{(\alpha_R,n_R,S_R)}_{(\alpha_L,n_L,S_L),(n_P,S_P)}}^2}$. 
    \item Now using the fusion rules from \cref{sec:cg_restraints}, we perform state preparations to create the reduced blocks corresponding to each fusion channel, which depends on the value of the physical register.
    \begin{enumerate}
        \item $(n_P, S_P)=(1, 1/2), \ket{\mathrm{phys}}=\ket{kl} =\ket{10}$ \\ 
            \begin{enumerate}
                \item Using a single CNOT, store $k \oplus l$ on the first qubit of the physical register, making it $\ket{(k\oplus l)l}$. Note, since the physical register in the $\ket{0}_{\text{reflect}}$ branch is always $\ket{00}$ we do not actually require to control on the reflection ancilla for this state preparation, only on the qubit storing the sum $k \oplus l$.
                \item Next, we need to update the $\rchi$ register utilising the fusion rules $n_L=n_R-1$, $S_L=S_R-1/2$. Controlled on $k \oplus l$, we simply flip the register using w(XOR($n_R$,$n_R-1$)) and w(XOR($2S_R, 2S_R+1$)) CX gates where $w(\cdot)$ is the hamming weight. 
                \item Next, controlled on $k \oplus l$, we prepare a state over a temporary qubit, 
                \begin{align}
                    \eta_{\alpha_R,n_R,S_R-\frac{1}{2}} \ket{0,n_R-1,S_R-\frac{1}{2},m_R}\ket{0}_\text{temp}+\eta_{\alpha_R,n_R,S_R+\frac{1}{2}} \ket{0,n_R-1,S_R-\frac{1}{2},m_R}\ket{1}_\text{temp}
                \end{align}
                where
                $\eta_{\alpha_R,n_R,S_R \pm 1/2} = \sqrt{\sum_{(\alpha_L,n_R-1,S_R\pm\frac{1}{2})}|(A^{[i]})^{(\alpha_R,n_R,S_R)}_{(\alpha_L, n_R-1, S_R \pm 1/2)}|^2} / \gamma_{10}$
                . Note, at this stage $(n_R, S_R)$ are fixed, and hence we are summing over admissible $\alpha_L$. Controlled on the $\ket{1}_\mathrm{temp}$ we then correct the $S$ register to contain the value $S_R+1/2$ via w(XOR($2(S_R-1/2) +1,2(S_R+1/2)+1$)) CX gates. 
                \item Finally, controlled  on $k \oplus l$, and the value of temp, or just $k \oplus l \oplus \text{temp}$ for the first and temp for the second state preparation below, we implement
                \begin{align}
                    &\ket{0,n_R-1,S_R-\frac{1}{2},m_R}\ket{0}_\text{temp} \to \nonumber \\
                    &\frac{1}{\xi_{\alpha_R,n_R,S_R-\frac{1}{2}}}\sum_{\alpha_L}(A^{[i]})^{(\alpha_R,n_R,S_R)}_{(\alpha_L,n_R-1,S_R-\frac{1}{2}),(1,\frac{1}{2})}\ket{\alpha_L,n_R-1,S_R-\frac{1}{2},m_R}\ket{0}_\text{temp} \ , \\ 
                    &\ket{0,n_R-1,S_R+\frac{1}{2},m_R}\ket{1}_\text{temp} \to \nonumber \\
                    &\frac{1}{\xi_{\alpha_R,n_R,S_R+\frac{1}{2}}}\sum_{\alpha_L}(A^{[i]})^{(\alpha_R,n_R,S_R)}_{(\alpha_L,n_R-1,S_R+\frac{1}{2}),(1,\frac{1}{2})}\ket{\alpha_L,n_R-1,S_R+\frac{1}{2},m_R}\ket{1}_\text{temp} ,
                \end{align}
            where the normalization factor 
            $$\xi_{\alpha_R,n_R,S_R\pm\frac{1}{2}}=\sqrt{\sum_{(\alpha_L,n_R-1,S_R\pm\frac{1}{2})}|(A^{[i]})^{(\alpha_R,n_R,S_R)}_{(\alpha_L,n_R-1,S_R\pm\frac{1}{2}),(n_P,S_P)}|^2}$$
            is given by the numerator in $\eta$'s expression.
            \item Disentangle temp register using the distinguishing bit of $2S_R, 2S_R+2$ (discussed later) and $k\oplus l$ with a Toffoli and uncompute the sum $k\oplus l$.
            \end{enumerate}

        \item $(n_P, S_P) = (2, 0), \ket{phys}=\ket{11}$
            \begin{enumerate}
                \item We control on the second qubit of the physical register. For that same reasoning as previously, there is no reason to control on the one of the reflection ancilla here. 
                \item Again, we need to update the $\rchi$ register utilizing the fusion rules. This time the relevant rules are $n_L=n_R-2$, $S_L=S_R$. Controlled on $l$, we simply flip the register using w(XOR($n_R$, $n_R-2$)) CX gates..
                \item Finally, controlled on $l$ we perform the state preparation over the $\alpha_L$'s,
                \begin{align}
                 \ket{0,n_R-2,S_R,m_R}\ket{11}_i \to \frac{1}{\gamma_{11}}\sum_{\alpha_L}(A^{[i]})^{(\alpha_R,n_R,S_R)}_{(\alpha_L,n_R-2,S_R),(2,0)}\ket{\alpha_L,n_R-2,S_R,m_R}\ket{11}_i.  
                \end{align}
            \end{enumerate}
        \item ($n_P, S_p)=(0, 0), \ket{phys}=\ket{00}$
            \begin{enumerate}
                \item For this case we need to control on both the first qubit $k=0$ and the $\ket{1}_{reflect}$.
                \item The fusion rules are trivial here, that is, $n_R=n_L, S_R=S_L$. 
                \item So we just need to perform a doubled-controlled state preparation,
                \begin{align}
                    &\ket{0,n_R,S_R,m_R}\ket{00}_i \to \frac{1}{\gamma_{00}}\sum_{\alpha_L}(A^{[i]})^{(\alpha_R,n_R,S_R)}_{(\alpha_L,n_R,S_R),(0,0)}\ket{\alpha_L,n_R,S_R,m_R}\ket{00}_i \ .
                \end{align}
            \end{enumerate}
    \end{enumerate}
\end{enumerate}

\subsection{Finding the distinguishing bit of $S$ register}\label[appsec]{appssec:distbit}

The distinguishing bit has been mentioned a few times in the main text, as a way to lower the cost in certain controlled operations. 

In the reduced tensor preparation, we will need to distinguish the two non trivial sub-cases, $S_L=S_R\pm1/2$, that is  $2S_R, 2S_R+2$ in integerized form. Since we know the value of $S_R$ during loading, we can follow classical logic to find a single distinguishing bit. The relevant cases are:
    \begin{enumerate}
        \item We want the distinguishing bit of $2S_R$ and (i) $2S_R$ is odd or (ii) $2S_R+1$ is odd
        \item We want the distinguishing bit of $2S_R+2$ and (i) $2S_R+1$ is odd or (ii) $2S_R$ is odd
    \end{enumerate}
In the subcases 1(i) and 2(i), our distinguishing bit will be the location of the second LSB where we check for equality to the same bit in the comparison value, i.e. $2S_R, 2S_R+2$, respectively. In the subcase 1(ii), one can write the cases where $2S_R \pmod 4 \ 0, 2$, and in each case, it can be shown that an open/closed control flip, controlled on the second LSB with target on the LSB, produces the distinguishing bit. The subcase 2(ii) can be handled similarly.

Similar calculations are made for CG loading to find the distinguishing bit (hexagon) in \cref{fig:CG}.

\subsection{Alternative reduced tensor loading}\label[appsec]{appssec:altreducedtensorloading}

\begin{figure}
    \centering
    \includegraphics[width=1.0\linewidth]{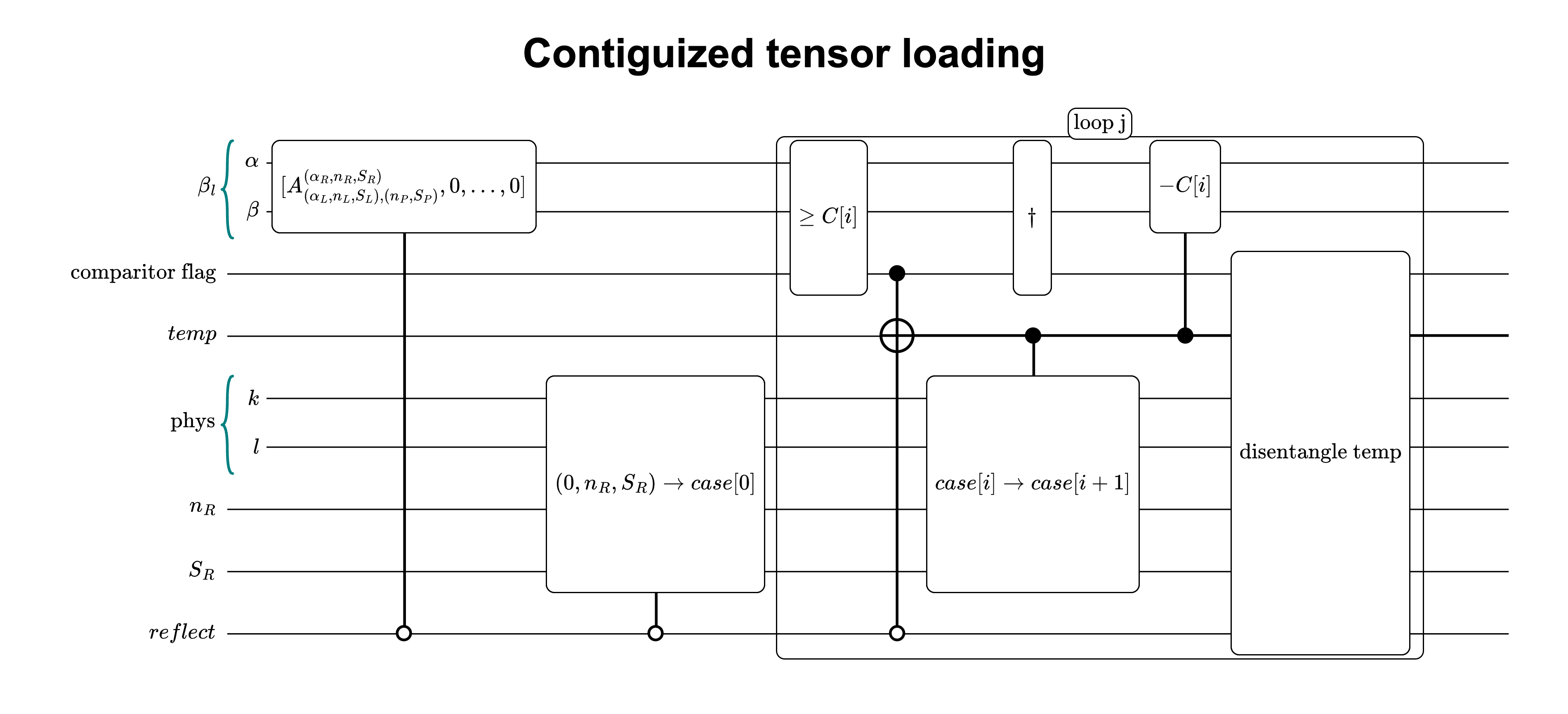}
    \caption{Contiguized tensor loading schematic. We first perform a state preparation on the extended $\beta_l$ register with ordering determined by multiplicity and appropriate padding. The list of `cases' are the correspondingly ordered $(n_L, S_L)(n_P, S_P)$. Note, here the $\dagger$-box corresponds to the uncomputation of the comparator.}
    \label{fig:alternativeloading}
\end{figure}
The circuit schematic is found in \cref{fig:alternativeloading}.
\begin{enumerate}
    \item We implement directly the full controlled state preparation
\begin{align}\label{eq:bigstatealt}
    \ket{1}_{\text{reflect}}\otimes \ket{0,n_R,S_R,m_R}_{\rchi} \to  \sum_{\beta_L \in \{(\alpha_L,n_L,S_L),(n_P,S_P)\}}(A^{[i]})^{(\alpha_R,n_R,S_R)}_{(\alpha_L,n_L,S_L),(n_P,S_P)}\ket{\beta_L,n_R,S_R,m_R}\ket{00}_i\ ,
\end{align}
    Note, that the summand is over the full column and is enumerated by $\beta_L$, which we call the contiguizer. This labels the left indices given by: $(n_L,S_L,n_P,S_P) = (n_R,S_R,0,0), (n_R-2,S_R,2,0),(n_R-1,S_R-1/2,1,1/2),(n_R-1,S_R+1/2,1,1/2)$. To implement this state prep, we first need to set a convention for the we contiguize) these sections of the reduced tensor. This convention is given by increasing order of the multiplicities of $(n_L,S_L)$ as they appear in $B_{n_R,S_R}$. From this ordering, we have an ordered array of multiplicities $\textbf{C}_L[0:4]$. It should be noted that not every block will include each of the four options enumerated by $\beta_L$, so the array in these cases will have less than four values. The size of the array will however always be classically known, with a maximum size of four. Due to there only being up to four possible $(n_L,S_L)$ options, and given that the size of the $\alpha$ register is the maximum log of all multiplicities, we only need a maximum of an additional two qubits added to the $\alpha$ register to store $\beta_L$. 
    \item \cref{eq:bigstatealt} needs an efficient decoding of the contiguizer $\beta_L$ and a correction on the occupation, spin, and physical register. \textit{Sequentially}, at each \textit{stage} $0\le j \le 3$, we compare the value of $\beta_L$ to the classically known value of $\text{C}_L[j]$ and store the corresponding flag $\beta_L \ge \text{C}_L[j]$ in a temporary ancilla. This ancilla is then used to perform the following flag-controlled operations:
    \begin{itemize}
        \item The values of the $n$, $S$, and physical registers are flipped from the $(j-1)$th case values $(n_L, S_L)(n_P, S_P)_{j-1}$ to the $j$th case values $(n_L, S_L)(n_P, S_P)_{j}$.
        \item The multiplicity $\textbf{C}_L[j]$ is subtracted from the $\beta_L$ register.
    \end{itemize}
    The flag ancilla is then disentangled. Note, all these operations are also controlled on the reflection ancilla. There is a slight distinction for the first case, with $j=0$, where the comparison and subtraction between the $\beta_L$ register and $\textbf{C}_L[j]$ is clearly unneeded. As a result here we simply write the 0th case values $(n_L, S_L)(n_P, S_P)_{j-1}$ controlled on the reflection ancilla.

    The total Toffoli cost from the above operations are three classically known comparators and controlled subtractors at each stage, each given by the dimension of the multiplicity register or twice that, respectively. The fact that one of the elements in the comparison/subtraction is classically known reduces the cost (for example by how many trailing zeros $\text{C}_L[j]$ has). The qubit cost is the size of the multiplicity register plus a few ancillae. This is all logarithmic and negligible in the total cost compared to the first state preparation and its (cleaned and reusable) qubit usage, esp. when performing qubit-gate tradeoff.
\end{enumerate}

\section{Resource estimation}\label[appsec]{appsec:resestimation}

\subsection{Toffoli count for LKS}\label[appsec]{appssec:tofflks}
Our resource estimation is done by Workbench, and for the controlled state preparation, it generally follows that of \cite[App. B]{berry2025rapid}, including its comment on combining the (largest) phase QROMs, but we have one adaptation and one change to the cited work.

First, our MPSs are all real, meaning all dense/reduced tensor amplitudes have real entries, and our largest phase QROM has output bit size $b=1$ to help determine the sign of the amplitudes, and not the default $b=15$ for arbitrary phase rotation. 

Second, in \cite[App. B]{berry2025rapid} Eq. (B1), the iterative part of LKS which is QROM + rotation + QROM$^\dagger$, has a derived cost
\begin{align}
    \bigg\lceil\frac{N}{2^{n-(k-1)}\Lambda}\bigg\rceil +\Lambda b+ \bigg\lceil\frac{N}{2^{n-(k-1)}\Lambda'}\bigg\rceil+ \Lambda'
    \label{eq:qromrotqromdaggercost_berry}
\end{align}
where $n=\lceil \log_2 N\rceil$. This cost is said to apply for all the iterative steps before the phase QROM, with the range of $k$ given as 3 to $n-1$, with the QROMs above acting on qubits 1 to $k-1$. However, if the largest $k$ is $n-1$, this means $k-1=n-2$, and thus the single qubit rotation followed by the QROMs produce a state superposition over $n-1$ qubits, and not $n$. What is missing is the last-to-one QROM, which reads $n-1$ qubits. Thus, the range of $k$ in the above should be $3$ to $n$. 

\subsection{Estimate qubit count for symmetric MPS loading}\label[appsec]{appsec:qubitcount}

\begin{table}[h!]
\centering
\begin{tabular}{ll}
\hline
Register & Qubit Count \\
\hline
System register & $2N$ \\
Virtual register & $\text{size}(\rchi)$ \\
QROM write register & $b$ \\
QROM & $\max_{L,\lambda} 1+(\lambda-1)b+\lceil \log(L/\lambda)\rceil$ \\
QROM$^\dagger$ & $\max_{L,\lambda}  1+\lambda+\lceil \log(L/\lambda)\rceil$\\
Adder & $b$ \\
Phase gradient & $b$ \\
Reflection ancilla & 1 \\
Temporary register  & 1  \\
(Subtractor/Comparator) & $\lceil\log_2(\max_{(n_i,S_i) \in \rchi_i,i} C_{n_i,S_i})\rceil$+2 \\
\hline
\end{tabular}
\caption{Qubit costings. The largest entries are the system register and the QROM's usage, the latter being larger, where one takes the maximum over all $L,\lambda$ instances appearing in the circuit. These qubits are cleaned and reused by other subroutines, such as the major ones like the subtractor/comparator in the alternative reduced tensor loading.}
\label{tab:qb_costs}
\end{table}

The main qubit costs for the SU(2) MPS circuit are summarized in \cref{tab:qb_costs}. The dense case is simpler and follows prior works in their baseline case \cite{rupprecht2026faster,berry2025rapid}.

For the QROM costings, the maximum of $L$ is $\max_{\rchi_i}C_{n_L,S_L}$, and for the alternative reduced tensor loading, $L = \max_{(n_R,S_R)\in \rchi_i,i}\text{sum}(\textbf{C}_L^{i})$, where the latter is the sum of the multiplicities of $(n_L,S_L)$ fusing into $(n_R,S_R)$. The  ancilla usage of the reduced tensor loading QROMs, esp. when trading off qubit for lower gate count, is dominant and can be reused by the comparator/subtractor circuits. 

We emphasize that our QREs are not symbolic like above. We used Workbench counters after compiling the entire circuit. The table above is to simply show where the main qubit costings come from, with an estimate of their count.

\section{Loading with direct unitary synthesis}\label[appsec]{appsec:direct}

In the construction of the main text, the restriction of the synthesized operator to the relevant Hilbert space is achieved with reflections. The Householder decomposition only needs the \emph{columns} $\ket{(T^{[i]})^{\rchi_i}}$ of \cref{eq:final_t_column}, and the reflection ancilla isolates them. The unitary-synthesis methods of \cite{berry2025rapid} and \cite{rupprecht2026faster} instead implement a unitary, or a prescribed set of columns of a unitary, directly. They use no reflection ancilla, and the synthesized operator acts on every basis state of the registers it touches. Using them for a U(1)~$\times$~SU(2) MPS therefore requires specifying:
\begin{enumerate}
    \item[(i)] which operator is synthesized and on which registers it acts;
    \item[(ii)] the subspace on which it must act correctly, and the freedom outside that subspace;
    \item[(iii)] how the right-sector labels $(n_R,S_R)$ that select a reduced block are used as controls, and how they are consistently transformed into $(n_L,S_L)$;
    \item[(iv)] how the Clebsch--Gordan coefficients are applied when no conjugation of reflections is available.
\end{enumerate}
This appendix gives a construction answering (i)--(iv). Its only data-dependent component is a \emph{sector-multiplexed reduced isometry}. That would be where one would simply plug in any direct unitary syntheses methods, e.g. those of \cite{berry2025rapid} and \cite{rupprecht2026faster}. 

\subsection{Registers, conventions and target}\label[appsec]{appssec:direct_conventions}

We use the registers of the main text.
\begin{itemize}
    \item The virtual register $\rchi$ consists of four parts. A multiplicity register $\alpha$ of $a=\lceil\log_2 C_{\max}\rceil$ qubits, where $C_{\max}:=\max_{i,(n,S)\in\rchi_i}C_{n,S}$, stores the multiplicity label in $0$-based form, $\alpha\in\{0,\dots,C_{n,S}-1\}$. A register $n$ of $r_n$ qubits stores the particle number. A register $S$ of $r_S$ qubits stores the integer $2S+1$. A register $m$ of $r_m$ qubits stores the integer $2m$ in two's complement. Any other affine encoding of $S$ and $m$ only changes the constants that are added below.
    \item The physical register of site $i$ consists of two qubits, $k$ (first) and $l$ (second), with the encoding of \cref{eq:P_i}. In particular, $\ket{10}_i$ and $\ket{01}_i$ carry $m_P=+\frac12$ and $m_P=-\frac12$, respectively.
    \item ``anc'' collects all ancillas of a subroutine: QROAM data registers, address and offset registers, flags, etc. (The phase-gradient register used for rotations is a catalyst and is returned unchanged).
\end{itemize}
We write $s=(n_R,S_R)\in\rchi_i$ for a right sector of site $i$, $C_s:=C^{[i]}_{n_R,S_R}$ for its multiplicity, and $K_i$ for the number of right sectors of site $i$.

The site circuit $U^{[i]}$ must satisfy, for all $(\alpha_R,n_R,S_R)\in\rchi_i$ and all $m_R\in\{-S_R,\dots,S_R\}$,
\begin{equation}\label{eq:direct_target}
    U^{[i]}\ket{\alpha_R,n_R,S_R,m_R}_{\rchi}\ket{00}_i\ket{0}_{\mathrm{anc}}=\ket{(T^{[i]})^{\rchi_i}}\ket{0}_{\mathrm{anc}},
\end{equation}
where $\ket{(T^{[i]})^{\rchi_i}}$ is given by \cref{eq:final_t_column}. The action of $U^{[i]}$ on any other input may be chosen freely, but must be completed to a unitary on the full registers.

\subsection{Fusion channels and the coupled channel code}\label[appsec]{appssec:channel_code}

\begin{dfn}[Fusion channels and channel code]\label{dfn:channels}
For a right sector $s=(n_R,S_R)$, the four fusion channels $j\in\{0,1,2,3\}$ are listed below, together with their left sectors $\lambda_j(s)=(n_L,S_L)$, physical labels $\pi_j=(n_P,S_P)$, coupling signs $\tau_j=2(S_R-S_L)$, and two-bit codes $\Gamma(j)=kl$.
\begin{center}
\begin{tabular}{ccccc}
\toprule
$j$ & $\lambda_j(s)=(n_L,S_L)$ & $\pi_j=(n_P,S_P)$ & $\tau_j$ & $\Gamma(j)$ \\
\midrule
$0$ & $(n_R,S_R)$ & $(0,0)$ & $0$ & $00$ \\
$1$ & $(n_R-2,S_R)$ & $(2,0)$ & $0$ & $11$ \\
$2$ & $(n_R-1,S_R-\frac12)$ & $(1,\frac12)$ & $+1$ & $10$ \\
$3$ & $(n_R-1,S_R+\frac12)$ & $(1,\frac12)$ & $-1$ & $01$ \\
\bottomrule
\end{tabular}
\end{center}
The channel multiplicity is $c_{s,j}:=C^{[i-1]}_{\lambda_j(s)}$ if $\lambda_j(s)\in\rchi_{i-1}$ (in particular $S_L\ge 0$), and $c_{s,j}:=0$ otherwise. We set $M_s:=\sum_j c_{s,j}$; this is $M_{n_R,S_R}$ of \cref{eq:dimmaxchi_i}, restricted to the sectors present in $\rchi_{i-1}$. The reduced block in channel form is the $M_s\times C_s$ matrix
\begin{equation}\label{eq:Bs_channel}
    (B_s)_{(j,\alpha_L),\alpha_R}:=(A^{[i]})^{(\alpha_R,n_R,S_R)}_{(\alpha_L,\lambda_j(s)),\pi_j},\qquad 0\le\alpha_L<c_{s,j},\quad 0\le\alpha_R<C_s.
\end{equation}
It is an isometry, $B_s^\dagger B_s=\mathrm{1}_{C_s}$, by \cref{eq:A_isometry}. We write $B_{s,j}$ for its $c_{s,j}\times C_s$ sub-block of channel $j$.
\end{dfn}

\begin{rmk}[What the code stores]\label{rmk:code}
Recall that $\ket{10}$ and $\ket{01}$ are the physical basis states with $m_P=\pm\frac12$. \textbf{\textit{Before}} the Clebsch--Gordan stage, however, they represent $\Gamma(2), \Gamma(3)$ and thus encode the coupling sign $\tau_j$, that is, whether $S_L=S_R-\frac12$ or $S_L=S_R+\frac12$, and not a spin projection. In the reduced-tensor loading of \cref{ssec:reducedtensor}, by contrast, both singly-occupied channels are written as $\ket{10}_i$ and the temp qubit holds $\tau$. Because $\Gamma$ is injective, the right sector is a function of the left sector and the code,
\begin{equation}\label{eq:code_inverse}
    n_R=n_L+k+l,\qquad 2S_R+1=(2S_L+1)+k-l,
\end{equation}
so the labels needed by the Clebsch--Gordan coefficients are never lost.
\end{rmk}

\subsection{The site circuit}\label[appsec]{appssec:four_stages}

\begin{figure}[t]
    \centering
    \includegraphics[width=\linewidth]{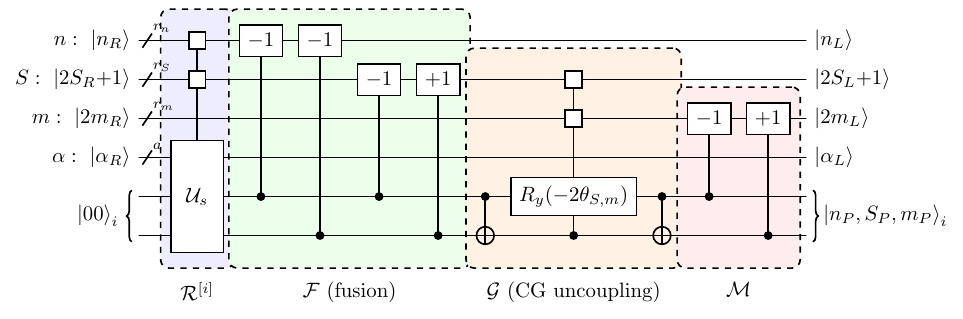}
    \caption{Site circuit for $U^{[i]}_{\mathrm{dir}}=\mathcal{M}\mathcal{G}\mathcal{F}\mathcal{R}^{[i]}$ of \cref{dfn:direct_site}. Squares denote multiplexer controls: the operation applied depends on the value held by the register.
    In $\mathcal{R}^{[i]}$, a unitary $\mathcal{U}_s$ that depends on the right sector $s=(n_R,S_R)$ acts on the multiplicity register and on the two physical qubits of site $i$ (top: $k$, bottom: $l$). It leaves the physical qubits in the coupled channel code $\Gamma(j)$ of \cref{dfn:channels}, and it does not modify $n$, $S$ or $m$.
    $\mathcal{F}$ applies the fusion rules with four singly-controlled $\pm1$ operations. $\mathcal{G}$ maps the coupled codes $\ket{10}$ and $\ket{01}$ (coupling signs $\tau=\pm1$) to the physical spin states with Clebsch--Gordan amplitudes, using a rotation multiplexed on $(S_L,m_R)$. $\mathcal{M}$ sets $m_L=m_R-m_P$.
    Only $\mathcal{R}^{[i]}$ depends on the MPS data. 
    }
    \label{fig:us_overview}
\end{figure}

\begin{dfn}\label{dfn:direct_site}
Let $U^{[i]}_{\mathrm{dir}}:=\mathcal{M}\,\mathcal{G}\,\mathcal{F}\,\mathcal{R}^{[i]}$ (\cref{fig:us_overview}), with the following four stages.
\begin{enumerate}
    \item The \emph{sector-multiplexed reduced isometry} is
    \begin{equation}\label{eq:R_def}
        \mathcal{R}^{[i]}=\sum_{s\in\rchi_i}\ketbra{s}{s}_{nS}\otimes \mathcal{U}_s\otimes\mathrm{1}_m+\Pi^{\perp}_{nS}\otimes X^{\perp}.
    \end{equation}
    Here $\Pi^\perp_{nS}$ projects onto the values of $(n,S)$ that are not right sectors of site $i$, and $X^\perp$ is an arbitrary unitary. $\mathcal{U}_s$ is a unitary on $(\alpha,k,l,\mathrm{anc})$ with
    \begin{equation}\label{eq:Us_def}
        \mathcal{U}_s\ket{\alpha_R}_\alpha\ket{00}_{kl}\ket{0}_{\mathrm{anc}}=\sum_{j=0}^{3}\sum_{\alpha_L=0}^{c_{s,j}-1}(B_s)_{(j,\alpha_L),\alpha_R}\ket{\alpha_L}_\alpha\ket{\Gamma(j)}_{kl}\ket{0}_{\mathrm{anc}},\qquad 0\le\alpha_R<C_s.
    \end{equation}
    The registers $n$ and $S$ are only read, as multiplexer controls, and $m$ is a spectator.
    \item The \emph{fusion update} $\mathcal{F}$ performs $n\leftarrow n-k-l$ and $(2S{+}1)\leftarrow(2S{+}1)-k+l$. It consists of two decrements of $n$ controlled by $k$ and $l$, and a decrement and increment of $S$ controlled by $k$ and $l$, respectively.
    \item The \emph{Clebsch--Gordan uncoupling} is
    \begin{equation}\label{eq:G_def}
        \mathcal{G}=\mathrm{CNOT}_{k\to l}\Big(\sum_{S,m}\ketbra{S,m}{S,m}\otimes\big(\mathrm{1}_k\otimes\ketbra{0}{0}_l+R_y(-2\theta_{S,m})_k\otimes\ketbra{1}{1}_l\big)\Big)\mathrm{CNOT}_{k\to l},
    \end{equation}
    where $R_y(\varphi)=e^{-i\varphi Y/2}$. For register contents $2S_L+1$ and $2m_R$, the angle is
    \begin{equation}\label{eq:theta_def}
        \theta_{S_L,m_R}:=\arccos\sqrt{\frac{S_L+m_R+\frac12}{2S_L+1}}\in\Big[0,\frac{\pi}{2}\Big],\qquad\text{equivalently}\quad\cos^2\theta_{S_L,m_R}=\frac{\mathfrak{s}+\mathfrak{m}}{2\mathfrak{s}},
    \end{equation}
    with $\mathfrak{s}=2S_L+1$ and $\mathfrak{m}=2m_R$. The formula is used only when $\mathfrak{s}\ge1$, $|\mathfrak{m}|\le\mathfrak{s}$ and $\mathfrak{s}+\mathfrak{m} = 2(S_L+S_R+1/2)$ is even. All other binary addresses, including $\mathfrak{s}=0$, are assigned angle zero. 
    \item The \emph{magnetic update} $\mathcal{M}$ performs $2m\leftarrow 2m-k+l$, that is, $m_L=m_R-m_P$.
\end{enumerate}
\end{dfn}

\begin{prop}\label{prop:G}
Fix $S_L\ge0$ and $m_R$ with $\abs{m_R}\le S_L+\frac12$ with $2S_L+1+2m_R \equiv 0 \pmod 2$, and let
\[
    c=\sqrt{\frac{S_L+m_R+\frac12}{2S_L+1}},\qquad \bar c=\sqrt{\frac{S_L-m_R+\frac12}{2S_L+1}}.
\]
On $\ket{2S_L+1,2m_R}_{Sm}\otimes\mathbb{C}^4_{kl}$, $\mathcal{G}$ acts as the identity on $\ket{00}$ and $\ket{11}$, and
\begin{equation}\label{eq:G_action}
    \ket{10}\mapsto c\ket{10}+\bar c\ket{01},\qquad\ket{01}\mapsto -\bar c\ket{10}+c\ket{01}.
\end{equation}
Consequently, for $j\in\{2,3\}$ and $S_R=S_L+\tau_j/2$,
\begin{equation}\label{eq:G_cg}
    \mathcal{G}\ket{2S_L{+}1,2m_R}\ket{\Gamma(j)}_i=\sum_{m_P=\pm\frac12}CG^{S_R,m_R}_{S_L,m_R-m_P;\frac12,m_P}\ket{2S_L{+}1,2m_R}\ket{1,\tfrac12,m_P}_i.
\end{equation}
\end{prop}
\begin{proof}
$\mathrm{CNOT}_{k\to l}$ maps $\ket{00},\ket{11},\ket{10},\ket{01}$ to $\ket{00},\ket{10},\ket{11},\ket{01}$, respectively. Between the two CNOTs, the singly-occupied states are therefore exactly those with $l=1$, with $k=1$ for the code $10$ and $k=0$ for the code $01$. The empty and doubly-occupied states have $l=0$ and are not rotated. With $\cos\theta=c$ and $\sin\theta=\bar c$, we have $R_y(-2\theta)\ket{1}=\bar c\ket{0}+c\ket{1}$ and $R_y(-2\theta)\ket{0}=c\ket{0}-\bar c\ket{1}$. The second CNOT maps $\ket{11}\mapsto\ket{10}$ and $\ket{01}\mapsto\ket{01}$, which gives \cref{eq:G_action}.
For $j=2$ we have $S_L=S_R-\frac12$, and \cref{eq:CGcoeffS_Rminushalf} gives $CG(m_P=+\frac12)=c$ and $CG(m_P=-\frac12)=\bar c$. For $j=3$ we have $S_L=S_R+\frac12$, and \cref{eq:CGcoeffS_Rplushalf} gives $CG(m_P=+\frac12)=-\bar c$ and $CG(m_P=-\frac12)=c$. With the encoding \cref{eq:P_i} of $m_P$, this is \cref{eq:G_cg}.
\end{proof}

\begin{thm}\label{thm:direct}
$U^{[i]}_{\mathrm{dir}}$ satisfies \cref{eq:direct_target}.
\end{thm}
\begin{proof}
Let $s=(n_R,S_R)$. By \cref{eq:R_def,eq:Us_def},
\[
    \mathcal{R}^{[i]}\ket{\alpha_R,n_R,S_R,m_R}\ket{00}_i\ket{0}=\sum_{j,\alpha_L}(B_s)_{(j,\alpha_L),\alpha_R}\ket{\alpha_L,n_R,S_R,m_R}\ket{\Gamma(j)}_i\ket{0}.
\]
On the branch of channel $j$, $\mathcal{F}$ replaces $(n_R,2S_R+1)$ by $(n_R-k-l,\,2S_R+1-k+l)$, which by the table of \cref{dfn:channels} is $\lambda_j(s)$.
For $j\in\{0,1\}$, $\mathcal{G}$ and $\mathcal{M}$ act trivially. This agrees with $CG^{S_R,m_R}_{S_L,m_L;0,0}=\delta_{m_L,m_R}$ and with $\ket{\Gamma(0)}=\ket{0,0,0}_i$ and $\ket{\Gamma(1)}=\ket{2,0,0}_i$.
For $j\in\{2,3\}$, \cref{prop:G} produces
\[
    \sum_{m_P=\pm\frac12}CG^{S_R,m_R}_{S_L,m_R-m_P;\frac12,m_P}\ket{\alpha_L,\lambda_j(s),m_R}\ket{1,\tfrac12,m_P}_i,
\]
and $\mathcal{M}$ replaces $m_R$ by $m_L=m_R-m_P$.
Summing over $j$, $\alpha_L$ and $m_P$ gives the second line of \cref{eq:final_t_column}, with every left label $(\alpha_L,n_L,S_L,m_L)$ and physical label $(n_P,S_P,m_P)$ appearing exactly once. The ancillas are returned to $\ket{0}$ by $\mathcal{R}^{[i]}$ 
and by $\mathcal{G}$, by definition.
\end{proof}

\begin{cor}\label[cor]{cor:mps}
Initialize the virtual register to $\ket{0,n_{\mathrm{tot}},2S_{\mathrm{tot}}+1,2m_{\mathrm{tot}}}$ for any $m_{\mathrm{tot}}\in\{-S_{\mathrm{tot}},\dots,S_{\mathrm{tot}}\}$, and all physical registers to $\ket{00}$. Then $U^{[1]}_{\mathrm{dir}}\cdots U^{[N]}_{\mathrm{dir}}$ prepares the SU(2) MPS with magnetization $m_{\mathrm{tot}}$ on the physical registers, and particle number and total spin $n_{\mathrm{tot}}, S_{\mathrm{tot}}$. It returns the virtual register to the vacuum $\ket{0,0,1,0}$ of $\rchi_0$ and all ancillas to $\ket{0}$.
\end{cor}

\section{Fermi--Hubbard MPS loading QREs}\label[appsec]{appsec:fermihubbardqre}
Below we list the numbers behind the FH QRE plots in \cref{sec:resourceestimation}. The last two columns Red. T and Red. AV  describe the Toffoli and active volume portion coming from the reduced tensor loading subroutine, which is itself dominated by the LKS.

\small
\begin{longtable}{r l r r r r r}
\toprule
$M$ & Circuit & Toffolis & Qubits & Active volume & Red. T & Red. AV \\
\midrule
\endhead
$2$ & Dense & $1{,}211{,}751$ & $89$ & $7.424\times10^{7}$ & \textemdash & \textemdash \\
$2$ & SU(2) & $43{,}276$ & $99$ & $3{,}457{,}611$ & $26.95\%$ & $45.03\%$ \\
$4$ & Dense & $1{,}220{,}993$ & $89$ & $7.494\times10^{7}$ & \textemdash & \textemdash \\
$4$ & SU(2) & $44{,}216$ & $100$ & $3{,}526{,}707$ & $27.13\%$ & $45.13\%$ \\
$8$ & Dense & $1{,}326{,}544$ & $89$ & $8.191\times10^{7}$ & \textemdash & \textemdash \\
$8$ & SU(2) & $50{,}774$ & $100$ & $4{,}091{,}849$ & $33.85\%$ & $50.83\%$ \\
$16$ & Dense & $1{,}847{,}601$ & $89$ & $1.146\times10^{8}$ & \textemdash & \textemdash \\
$16$ & SU(2) & $68{,}875$ & $101$ & $5{,}615{,}538$ & $46.88\%$ & $61.19\%$ \\
$32$ & Dense & $3{,}860{,}090$ & $90$ & $2.323\times10^{8}$ & \textemdash & \textemdash \\
$32$ & SU(2) & $116{,}459$ & $102$ & $9{,}363{,}034$ & $63.81\%$ & $73.43\%$ \\
$64$ & Dense & $8{,}327{,}298$ & $90$ & $5.030\times10^{8}$ & \textemdash & \textemdash \\
$64$ & SU(2) & $233{,}199$ & $102$ & $1.805\times10^{7}$ & $78.42\%$ & $83.67\%$ \\
$128$ & Dense & $2.709\times10^{7}$ & $91$ & $1.605\times10^{9}$ & \textemdash & \textemdash \\
$128$ & SU(2) & $565{,}901$ & $103$ & $4.156\times10^{7}$ & $87.98\%$ & $90.55\%$ \\
$256$ & Dense & $9.298\times10^{7}$ & $92$ & $5.423\times10^{9}$ & \textemdash & \textemdash \\
$256$ & SU(2) & $1{,}420{,}666$ & $104$ & $9.866\times10^{7}$ & $93.28\%$ & $94.49\%$ \\
$512$ & Dense & $3.061\times10^{8}$ & $93$ & $1.778\times10^{10}$ & \textemdash & \textemdash \\
$512$ & SU(2) & $3{,}720{,}489$ & $105$ & $2.489\times10^{8}$ & $96.02\%$ & $96.65\%$ \\
$1024$ & Dense & $1.073\times10^{9}$ & $94$ & $6.193\times10^{10}$ & \textemdash & \textemdash \\
$1024$ & SU(2) & $1.020\times10^{7}$ & $106$ & $6.550\times10^{8}$ & $97.74\%$ & $98.04\%$ \\
\bottomrule
\caption{$4\times4$, $n=0.875$ QREs for the dense and symmetric MPS loading up to bond dimension 1024. The policy used is $\lambda=1$, no qubit-gate trade-off. All numbers are from Workbench \cite{psiquantum_qdk} circuits, verified at small FH and molecular instances.}\label{tab:4x4fullfh}
\end{longtable}
\normalsize

\small
\begin{longtable}{r l r r r r r}
\toprule
$M$ & Circuit & Toffolis & Qubits & Active volume & Red. T & Red. AV \\
\midrule
\endhead
$2$ & Dense & $2{,}539{,}378$ & $105$ & $1.551\times10^{8}$ & \textemdash & \textemdash \\
$2$ & SU(2) & $78{,}019$ & $116$ & $6{,}308{,}123$ & $28.91\%$ & $47.63\%$ \\
$4$ & Dense & $2{,}542{,}913$ & $105$ & $1.552\times10^{8}$ & \textemdash & \textemdash \\
$4$ & SU(2) & $79{,}035$ & $117$ & $6{,}369{,}851$ & $28.88\%$ & $47.52\%$ \\
$8$ & Dense & $2{,}700{,}160$ & $105$ & $1.657\times10^{8}$ & \textemdash & \textemdash \\
$8$ & SU(2) & $88{,}341$ & $117$ & $7{,}145{,}926$ & $34.10\%$ & $51.61\%$ \\
$16$ & Dense & $3{,}404{,}801$ & $105$ & $2.103\times10^{8}$ & \textemdash & \textemdash \\
$16$ & SU(2) & $114{,}343$ & $118$ & $9{,}337{,}920$ & $44.86\%$ & $60.09\%$ \\
$32$ & Dense & $6{,}635{,}733$ & $106$ & $4.030\times10^{8}$ & \textemdash & \textemdash \\
$32$ & SU(2) & $196{,}671$ & $119$ & $1.589\times10^{7}$ & $62.66\%$ & $72.91\%$ \\
$64$ & Dense & $1.542\times10^{7}$ & $106$ & $9.338\times10^{8}$ & \textemdash & \textemdash \\
$64$ & SU(2) & $419{,}057$ & $119$ & $3.245\times10^{7}$ & $78.43\%$ & $83.82\%$ \\
$128$ & Dense & $5.213\times10^{7}$ & $107$ & $3.087\times10^{9}$ & \textemdash & \textemdash \\
$128$ & SU(2) & $1{,}080{,}879$ & $120$ & $7.884\times10^{7}$ & $88.53\%$ & $90.99\%$ \\
$256$ & Dense & $1.891\times10^{8}$ & $108$ & $1.105\times10^{10}$ & \textemdash & \textemdash \\
$256$ & SU(2) & $2{,}998{,}660$ & $121$ & $2.062\times10^{8}$ & $93.79\%$ & $94.90\%$ \\
$512$ & Dense & $6.916\times10^{8}$ & $109$ & $4.025\times10^{10}$ & \textemdash & \textemdash \\
$512$ & SU(2) & $8{,}799{,}346$ & $122$ & $5.792\times10^{8}$ & $96.53\%$ & $97.06\%$ \\
$1024$ & Dense & $2.659\times10^{9}$ & $110$ & $1.544\times10^{11}$ & \textemdash & \textemdash \\
$1024$ & SU(2) & $2.626\times10^{7}$ & $123$ & $1.676\times10^{9}$ & $98.00\%$ & $98.27\%$ \\
\bottomrule
\caption{$4\times6$, $n=0.75$ QREs for the dense and symmetric MPS loading up to bond dimension 1024. The policy used is $\lambda=1$, no qubit-gate trade-off. All numbers are from Workbench \cite{psiquantum_qdk} circuits, verified at small FH and molecular instances.}\label{tab:4x6fullfh}
\end{longtable}

\section{Fermi--Hubbard fidelity data}\label[appsec]{appsec:fermihubbard}
\begin{figure}[h!]
    \centering
    \includegraphics[width=0.35\columnwidth]{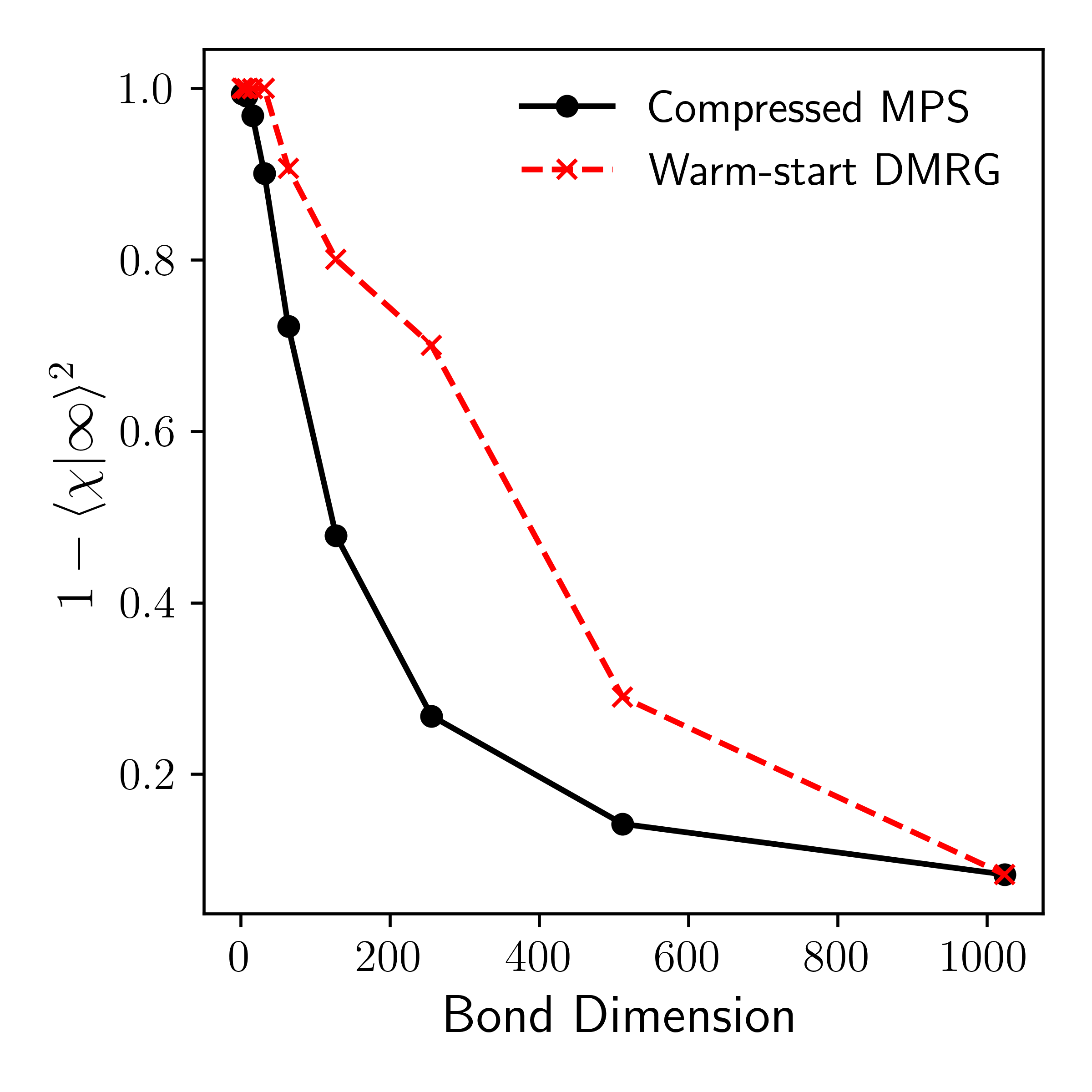}
    \caption{Comparison of initial-state infidelities for generating a MPS from warm-start DMRG versus compression from $\rchi_{\mathrm{SU(2)}} = 1024$ for the $4 \times 6$ Fermi--Hubbard lattice $n = 0.75$.}
    \label{fig:compress-vs-sweep-fidelity}
\end{figure}

For each lattice and filling, we first converge an SU(2)-symmetric MPS with bond dimension $\rchi_{\mathrm{SU(2)}}=1,024$ using DMRG. We then directly compress this state, without additional DMRG sweeps, to bond dimensions
$\rchi_{\mathrm{SU(2)}} = 2,4,8,16,32,64,128,256,$ and $512$, following the procedure of \cite{fomichev2024initial}. For both lattice sizes, we use a bond-dimension-$10{,}000$ MPS as a proxy for the exact ground state.

As an example, \cref{fig:compress-vs-sweep-fidelity} compares the fidelity between a warm-start DMRG calculation at a given bond-dimension (starting from the $1,024$ bond dimension MPS) and this compression technique for the $4 \times 6$ lattice with a particle/hole ratio of $n=0.75$. It is seen for every bond dimension, the computationally simpler compression approach has a lower infidelity as compared to variationally optimizing that state yielding a better quality initial state for a given bond dimension. The conclusion that this yields favorable initial states is in agreement with \cite{fomichev2024initial} for chemistry models. 

\begin{figure}[h!]
    \centering
    \includegraphics[width=0.35\columnwidth]{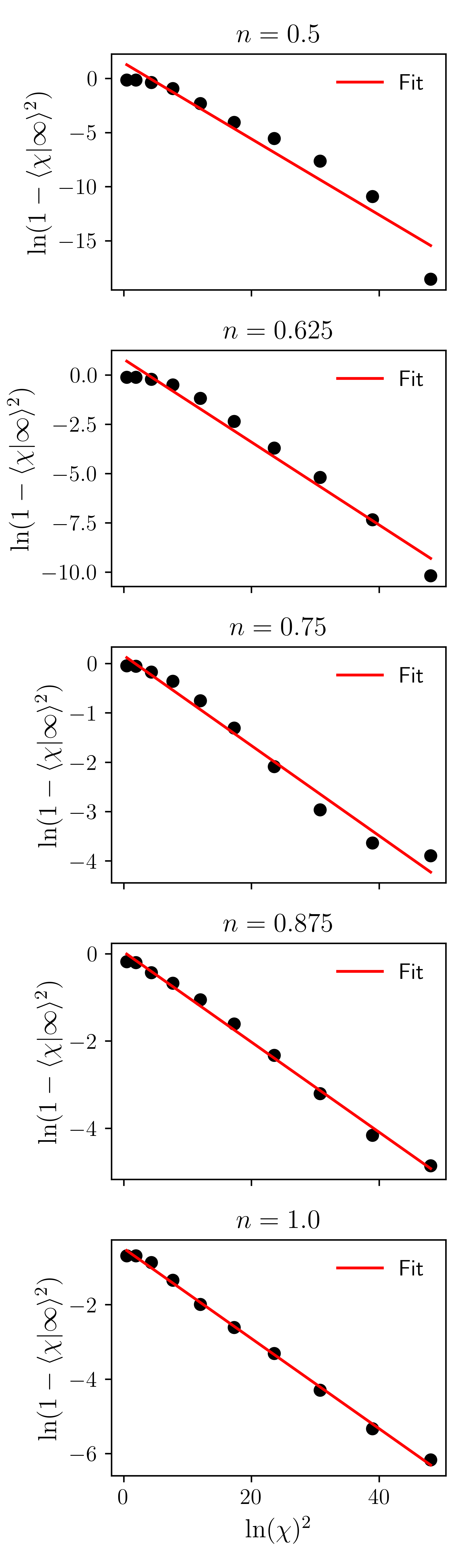}
    \includegraphics[width=0.35\columnwidth]{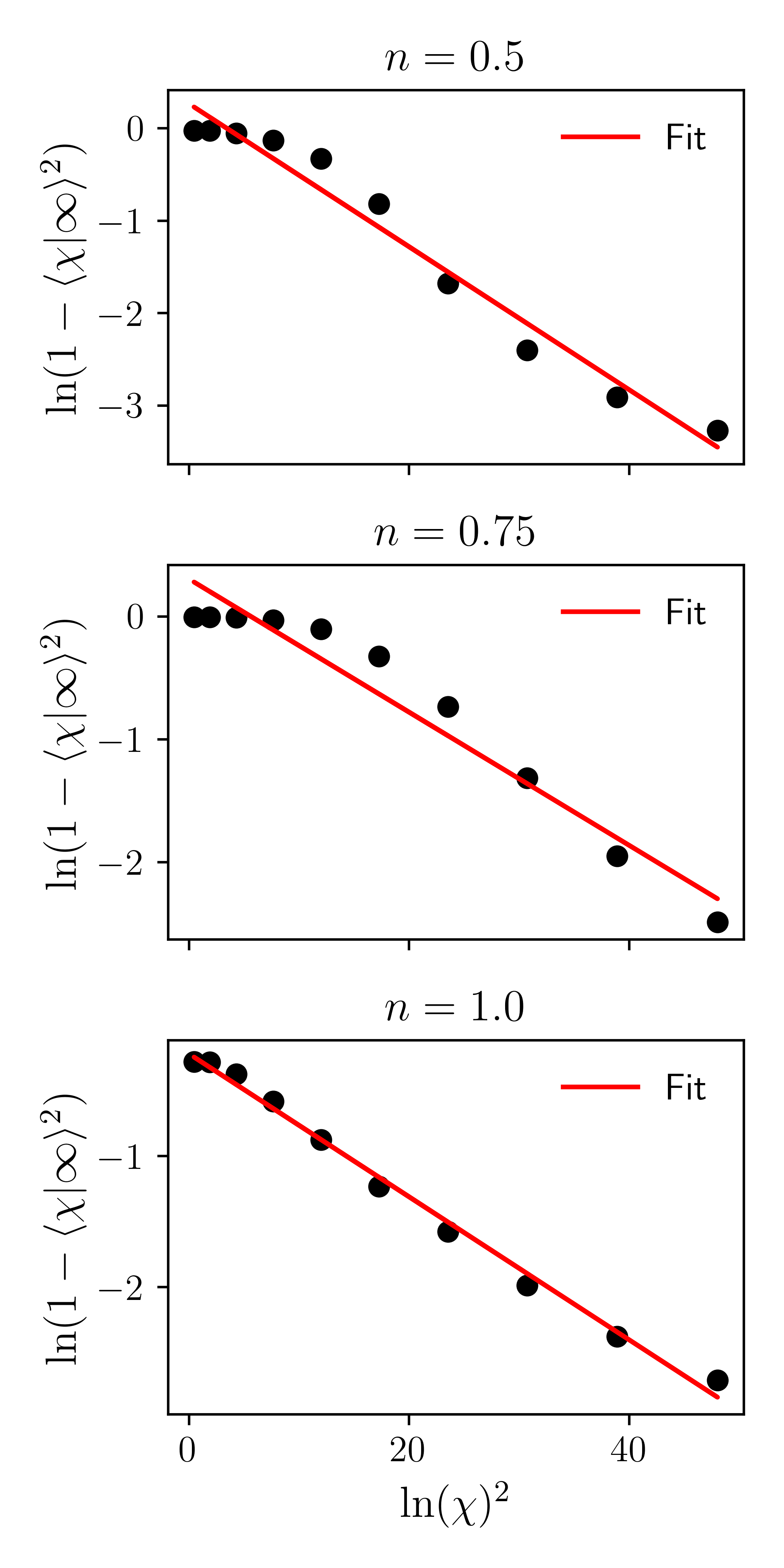}
    \caption{Infidelity for the $4\times 4$ and $4\times 6$ Fermi--Hubbard lattice with $U/t=8$ at and away from half-filling versus bond dimension. Linear regression of the data is shown in red confirming the empirical linear relationship between $\log(\rchi)^2$ and $\log(1-\langle \rchi | \infty\rangle ^2)$. $\ket{\infty}$ approximated by $\ket{10,000}$. $n$ is the particle/hole ratio.}
    \label{fig:4x44x6-linear-fidelity}
\end{figure}

\subsection{Estimating fidelity without the true ground state}

Knowing the fidelity of the initial state that is loaded is also important for accurate QRE. Though, calculating the fidelity of an arbitrary initial state typically requires also knowing the true ground-state wave function (or an extremely close approximation). However, this will become quickly classically intractable, especially as quantum computers are best posed to target problems that are classically hard. Instead, one must turn to fidelity estimation techniques to avoid explicitly calculating/estimating the true ground-state wave function. Recently, an empirically motivated extrapolation technique using the bond dimension was introduced where linear relationships between $\ln\pqty{1 - \braket{\rchi}{\infty} ^2}$ and $\ln\pqty{\rchi}^2$ and $\ln\pqty{\braket{\rchi}{\rchi^\prime}^2-\braket{\rchi}{\infty}^2}$ and $\ln\pqty{\rchi^\prime}^2$ can be utilized to estimate $\braket{\rchi}{\infty}^2$ for a given bond dimension MPS ($\rchi$) \cite{berry2025rapid}.

We investigated if the linear relationship between $\ln\pqty{1-\braket{\rchi}{\infty} ^2}$ and $\ln\pqty{\rchi}^2$ still held true for Fermi--Hubbard MPS initial states. \Cref{fig:4x44x6-linear-fidelity} shows the plots of $\ln\pqty{1-\braket{\rchi}{\infty}^2}$ versus $\ln\pqty{\rchi}^2$ for the $4 \times 4$ and $4 \times 6$ Fermi--Hubbard lattice with $U/t=8$ and various fillings. Our results largely confirm that the linear relationship holds for other lattices and fillings. We see that in all cases, we retain a nice linear relationship between the two quantities; however, there is some deviation from linearity at small bond dimensions ($\rchi < 10$). Here, we have focused on taking bond dimensions which range from 2 to 1024; Note that MPSs with sufficiently small bond dimension may not provide much meaningful information. In fact, bond dimensions are a unitless quantity and a bond dimension 16 can be quite large for some lattice sizes (it corresponds to exact diagonalization for the $2\times 2$ case), but small in other lattices ($4\times 4$ and beyond). In the future, we may explore which bond dimensions to retain for the linear extrapolation technique by renormalizing the bond dimension quantity so that it is relative to the problem instance. Alternatively, we may also explore other MPS metrics such as discarded weight instead of bond dimension as they are more physically meaningful in reporting how close an MPS is to exact solution.

Going to larger lattice sizes/higher bond dimensions takes weeks (to months) even if using more sophisticated and supervised PEPS techniques \cite{liu2025accuratesimulationhubbardmodel}. In essence, a DMRG-based initial state loading approach has substantial scaling problems on both the classical and quantum side of the cost equation.

\begin{table}
    \centering
    \begin{tabular}{llrr}
        \hline\hline
         & & \multicolumn{2}{c}{$\rchi_{\mathrm{dense}}/\rchi_{\mathrm{SU(2)}}$} \\
         \cline{3-4}
        $\rchi_{\mathrm{SU(2)}}$ && $4\times 4$ & $4\times 6$ \\
        \hline
                           2 &&    2.000000 &    2.000000 \\
                           4 &&    2.000000 &    2.000000 \\
                           8 &&    2.000000 &    1.875000 \\
                          16 &&    2.750000 &    2.750000 \\
                          32 &&    2.812500 &    2.937500 \\
                          64 &&    2.875000 &    2.828125 \\
                         128 &&    2.875000 &    2.906250 \\
                         256 &&    2.980469 &    3.085938 \\
                         512 &&    3.076172 &    3.197266 \\
                        1000 &&    2.994000 &    3.267578 \\
                       10000 &&    3.136200 &    3.492300 \\
        \hline\hline
    \end{tabular}
    \caption{\label{tab:bond_dim_ratio} Bond dimension ratio of dense MPS to SU(2) MPS loading computed from data. As expected, we observe around 2 and 3 times difference.}
    
\end{table}

\section{Density definition}\label[appsec]{appsec:densitydefinition}

Let us make a remark on the definition of density/sparsity. $\rho^{-1}$ in prior work is given in terms of the \textit{densified} U(1) MPS, as ``the quotient of the dense tensor sizes and the number of non-zero elements in the [dense] tensors'', which in \cref{tab:chemical_mps_loading_prior}, we have calculated it for our MPSs as ``our $\rho^{-1}$''. \cite{rupprecht2026faster} uses the dense MPS in both the numerator and denominator of their density definition, because they use a new unitary synthesis method applied on the dense MPS, which is always a densified symmetric MPS, which is the U(1) MPS in that work (the input behind their \textit{sparse} QRE entry). In general, considering the unitary synthesis algorithms (whether LKS or those of \cite{berry2025rapid,rupprecht2026faster}), in order for a `density' measure to be a target/representative for the expected speedup, the measure needs to be based on a ratio of the number of nonzero datapoints \textit{or} the size of the tensors in the algorithm's \textit{target}, be it dense or symmetric MPS. The dense \textit{and} symmetric MPS loading algorithm we used, are both based on the Householder-LKS unitary synthesis method, and they scale with the \textit{size} of the dense and reduced isometry tensors, respectively. Therefore, our measure of density should be a ratio of a quantity based on the densified symmetric MPS over a quantity based on the symmetric MPS, and not both based on the densified MPS; and this is how we have defined $s_{\mathrm{SU(2)}}^{-1}$. Though, we defined it based on the size of the dense tensors and the number of nonzero elements in the reduced tensors; the latter seems to be more relevant if we had actually used the sparse unitary synthesis in \cite{rupprecht2026faster} to load the reduced tensors directly, and not the Householder-LKS method as we did. However, as was the case for the lattice Hamiltonian benchmarks, the number of nonzero elements and the size of the reduced blocks are indeed close (32,729,316 and 32,736,934 for P450 ($S=0.5$) for a $10^{-8}$ threshold), meaning the reduced blocks are also \textit{dense} in these benchmarks with relatively few nonzero elements. So $s_{\mathrm{SU(2)}}^{-1}$, and not $\rho^{-1}$, is a reasonable target speedup in how we compare our own algorithms for dense and symmetric MPS loading. But that is literally the case when we are not trading off qubits with gates, as then the true speedup ratio given by the cost equations scales as something between $s_{\mathrm{SU(2)}}^{-1}$ and $\sqrt{s_{\mathrm{SU(2)}}^{-1}}$. This comparison then becomes useful to illustrate how much the underlying algorithms are suffering from overheads irrelevant to the number of (nonzero) datapoints, such as subroutines solely having to do with encodings or unfolding the symmetry (CG-coefficients), as well as the overheads of state preparation causing its cost to not scale exactly as the number of (nonzero) amplitudes.

\section{Fidelities for SU(2) MPS}\label[appsec]{appsec:fidelitiesavtables}

\Cref{tab:mps-fidelity-various-lattices} herein include fidelity of the DMRG data.

\begin{table}[h]
    \centering
    \resizebox{\textwidth}{!}{%
    \begin{tabular}{l rrrrr r rrr}
        \hline\hline
         & \multicolumn{5}{c}{$4\times 4$} & & \multicolumn{3}{c}{$4\times 6$} \\
         \cline{2-6}\cline{8-10}
        $\rchi_{\mathrm{SU(2)}}$  & $n=1.000$ & $n=0.875$ & $n=0.750$ & $n=0.625$ & $n=0.500$ && $n=1.00$ & $n=0.75$ & $n=0.50$ \\
        \hline
        2    &      0.50 &      0.10 &      0.04 &      0.16 &      0.13 && 0.25 & 0.01 & 0.02 \\
        4    &      0.50 &      0.11 &      0.05 &      0.18 &      0.13 && 0.25 & 0.01 & 0.03 \\
        8    &      0.58 &      0.20 &      0.16 &      0.35 &      0.31 && 0.31 & 0.01 & 0.05 \\
        16   &      0.74 &      0.39 &      0.30 &      0.49 &      0.61 && 0.44 & 0.03 & 0.12 \\
        32   &      0.86 &      0.91 &      0.53 &      0.65 &      0.90 && 0.58 & 0.10 & 0.28 \\
        64   &      0.93 &      0.98 &      0.73 &      0.80 &      0.98 && 0.71 & 0.28 & 0.56 \\
        128  &      0.96 &      0.99 &      0.88 &      0.90 &      0.99 && 0.79 & 0.52 & 0.81 \\
        256  &      0.99 &      1.00 &      0.95 &      0.96 &      1.00 && 0.86 & 0.73 & 0.91 \\
        512  &      1.00 &      1.00 &      0.97 &      0.98 &      1.00 && 0.91 & 0.86 & 0.95 \\
        1024 &      1.00 &      1.00 &      0.98 &      0.99 &      1.00 && 0.93 & 0.92 & 0.96 \\
        \hline\hline
    \end{tabular}}
    \caption{\label{tab:mps-fidelity-various-lattices} Overlap of the compressed MPS's with bond dimension $10,000$ MPS state (approximate for exact eigenstate) for various lattice sizes and fillings. $\rchi_{\mathrm{SU(2)}}$ corresponds SU(2) bond dimension.}
\end{table}

\section{Illustration of tensor block structure}
\label[appsec]{appsec:ill_block}

\begin{figure}[h]
    \centering
    \begin{subfigure}[t]{0.96\textwidth}
\centering
\includegraphics[width=0.75\linewidth]{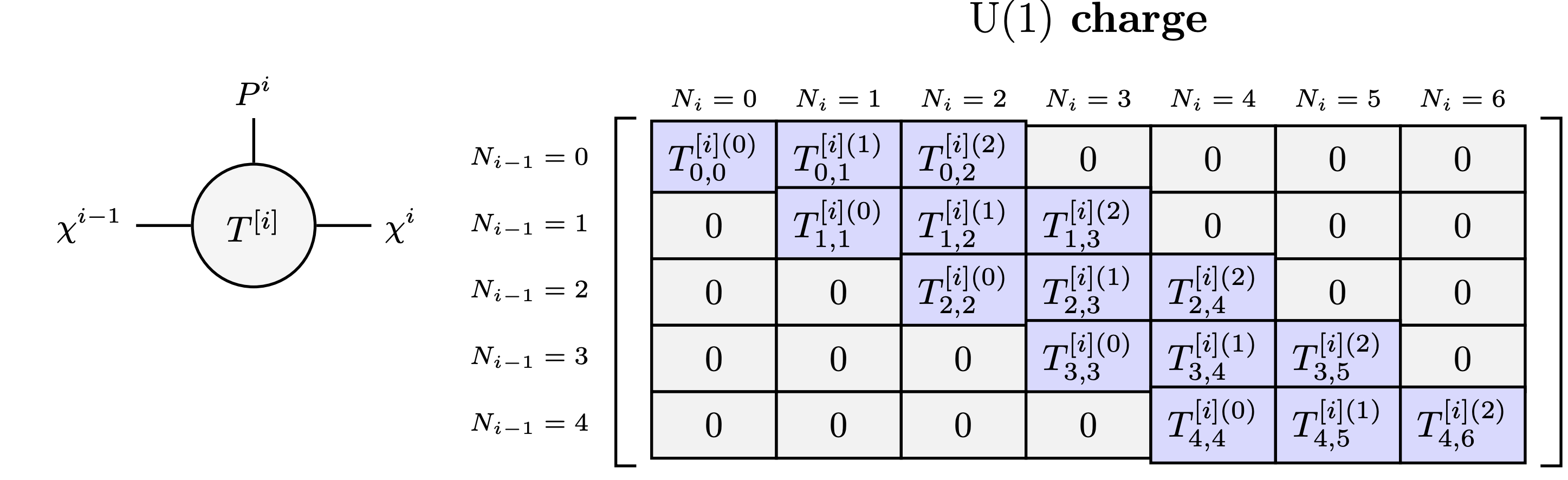}
    \caption{
U(1) particle-number selection rule for the site tensor
$T^{[i]}$. The local fermion basis
$P^i$ has charges $N_P=0,1,1,2$, and nonzero blocks satisfy
$N_i=N_{i-1}+N_P$. Beware of the change in notation; The superscript $(N_P)$ labels the physical charge sector, while the subscripts label (distinct) virtual charge sectors.
}
\label{fig:fermion-u1}
\end{subfigure}

\vspace{1.0em}

\begin{subfigure}[t]{0.96\textwidth}
\centering
\includegraphics[width=0.75\linewidth]{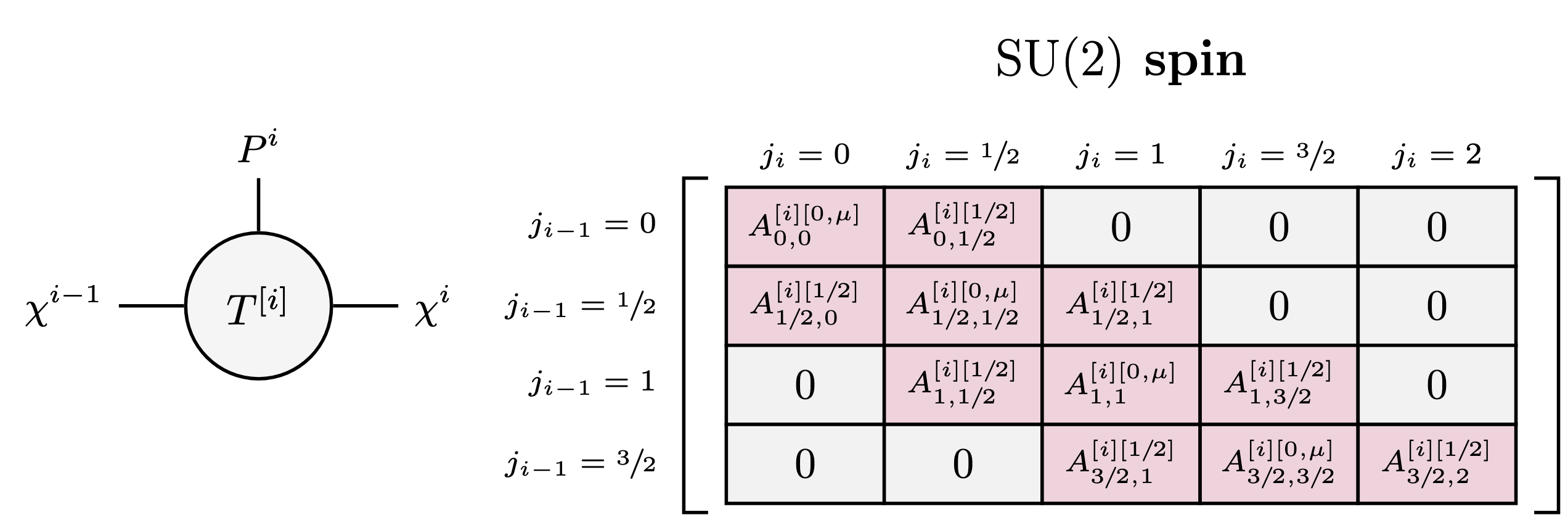}
\caption{
Spin SU(2) reduced-tensor illustration of $T^{[i]}$. The states
$P^i=0,2$ are spin singlets, denoted $j_P=0$ with charge
$\mu\in\{0,2\}$. Nonzero reduced blocks satisfy the spin fusion rules $j_i\in j_{i-1}\otimes j_P$. 
}
\label{fig:fermion-su2}
\end{subfigure}
\caption{
Symmetry-sector structure of a fermionic MPS tensor
$(T^{[i]})^{\rchi_i}_{\rchi_{i-1},P^i}$, with physical index
$P^i$ and virtual indices
$\rchi_{i-1},\rchi_i$. The panels show sector-level block structure rather than
individual virtual-index values; each sector may contain an arbitrary degeneracy space.
}
\label{fig:mps-sym}
\end{figure}

We present in \cref{fig:mps-sym} two illustrations of the block structures of a U(1) and SU(2) symmetric tensor of a fermionic MPS fermionic MPS. The tensor is given as $(T^{[i]})^{\rchi_i}_{\rchi_{i-1},P^i}$, where $P^i$ is the local physical index and $\rchi_{i-1},\rchi_i$ are the left and right virtual indices. 

\Cref{fig:fermion-u1} shows a fermionic U(1) symmetric tensors, where the leg charges determine which of the $P^{i}$ elements are non-zero. Often this is represented as a series of block diagonal matrices, especially if sequentially enumerated by $P^i$. Thus the tensor is block sparse, but the allowed blocks need not lie on the diagonal given the ordering of $P^i$.

\Cref{fig:fermion-su2} on the other hand shows an approximation of the SU(2) structure of the tensor. Unlike the U(1) case, we do not display sub-blocks of the tensor but instead the reduced tensor $A$ given from (schematically) $T^{[i]} =  A \otimes CG$ where $CG$ are the CG coefficients (see \cref{eq:TAC} for the full notation). We expect more non-zero blocks than the U(1) with additional minimal sizes according to the size of the CG coefficients.  
When considering the basis, the empty and doubly occupied states form spin singlets, while the singly occupied states form one spin-$\frac12$ multiplet. The displayed tensor is therefore a reduced tensor $A$ whose nonzero blocks obey the fusion rule
$j_i\in j_{i-1}\otimes j_P$. Due to the non-abelian nature of these tensors, there is additional constraints in its representation, namely the CG coefficient tensor that must be calculated and contracted appropriately within each tensor. These are omitted for brevity but explained in the main text.

\section{Hartree-Fock as an alternative: A small demo}
\label{sec:hfdemo}

\begin{figure}[h!]
    \centering
    \includegraphics[width=0.7\linewidth]{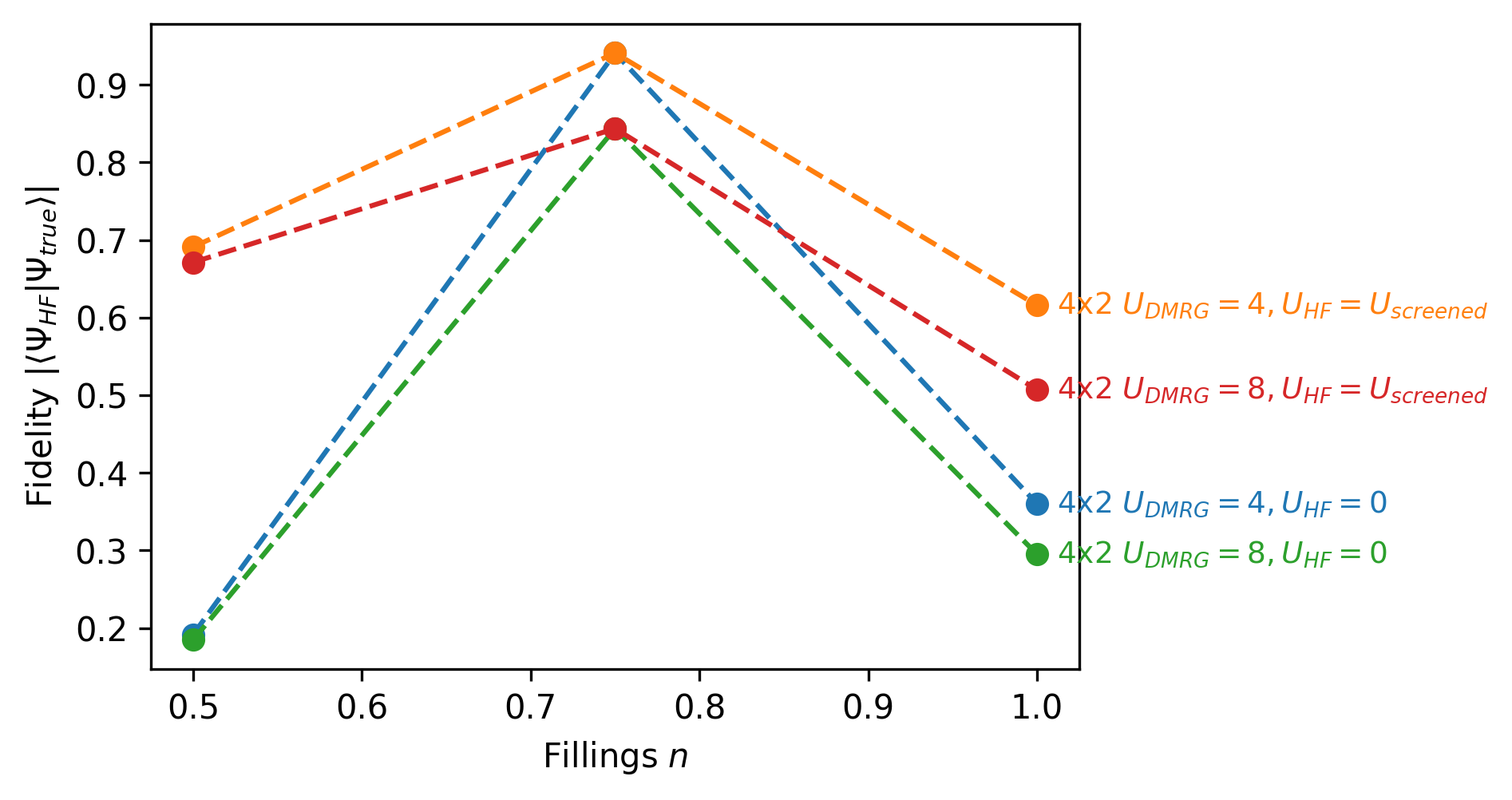}
    \caption{Preliminary absolute ground-state overlaps $|\langle\Psi_{\mathrm{HF}}|\Psi_{\mathrm{DMRG}}\rangle|$ for single-determinant trial states on the $4\times2$ Hubbard lattice. The DMRG state is obtained at sufficiently large bond dimension with negligible discarded weight and is used as the converged ground-state reference. The noninteracting determinant ($U_{\mathrm{HF}}=0$) is compared with a screened-HF determinant constructed using $U_{\mathrm{HF}}\simeq2.3$ for the target $U=4$ model and $U_{\mathrm{HF}}\simeq2.8$ for the target $U=8$ model. Lines are guides to the eye.}
    \label{fig:hf_screened_fidelity}
\end{figure}

Hartree-Fock (HF) is among the simplest initial state preparation methods, which provides a single Slater determinant as the lowest-energy state. This method easily scales up to more than 10,000-site systems, but the result usually overlaps too little with the true ground state. One way to increase the fidelity of the HF state is to use the screened Coulomb interaction to include the screening effect in HF calculations. 
\cref{fig:hf_screened_fidelity} benchmarks the fidelity of the screened HF method against the non-interacting state. The screened Coulomb potential helps raise the fidelity to larger than 50\% in all data points shown in the preliminary plot. 

The improvement is particularly pronounced at half filling and at $n_{\text{filling}}=0.5$, whereas the smaller change near $n_{\text{filling}}=0.75$ suggests that the optimal degree of screening depends on filling. These preliminary results illustrate that even very compact, efficiently preparable states can retain substantial ground-state weight when their classical construction is tailored to the interacting problem. Of course, the next step is to benchmark the HF fidelity against the system size after we converge the DMRG for more systems. The fidelity of HF state will likely decay according to the system size, but smarter post-processing such as symmetry-projection may still achieve a reasonable lower-bound and/or a coarse-QPE friendly spectrum.

\section{QPE emulation SU(2) versus $S_z$ MPS}
\label{sec:qpe_emulation}

To illustrate how spin adaptation affects a downstream quantum algorithm, we emulate quantum phase estimation (QPE) using low-bond-dimension DMRG states as the initial state. We consider the half-filled $2\times2$ Fermi--Hubbard model with $N_\uparrow=N_\downarrow=2$, $t=1$, and $U=8$.  Periodic boundaries follow the geometric convention used in other DMRG benchmarks. Two independent 20-sweep DMRG calculations were performed with a maximum symmetry-aware bond dimension $M=2$: one in the $N=4$, $S=0$ SU(2) sector and one in the $N=4$, $S_z=0$ abelian sector.  Here $M=2$ counts reduced multiplets in the SU(2) calculation but individual abelian states in the $S_z$ calculation, so equal nominal bond dimensions do not imply equal variational capacity.

Exact diagonalization (ED) gives all 36 eigenstates in the matching $N_\uparrow=N_\downarrow=2$ subspace. For each MPS $\ket{\psi}$ we compute
\begin{equation}
    c_n=\langle E_n|\psi\rangle,\qquad
    P(k)=\sum_{n=0}^{35}|c_n|^2 P_{\mathrm{sin}}^{(t_q)}(k\mid\phi_n),
    \label{eq:qpe_mixture}
\end{equation}
where $P_{\mathrm{sin}}^{(t_q)}$ is the finite-resolution QPE distribution for a sine-windowed phase register.  The ED energies are mapped into the phase interval by normalization of the spectrum.  We use $t_q=4$ phase qubits and 80 QPE shots.  

\begin{figure}[h!]
    \centering
    \includegraphics[width=\linewidth]{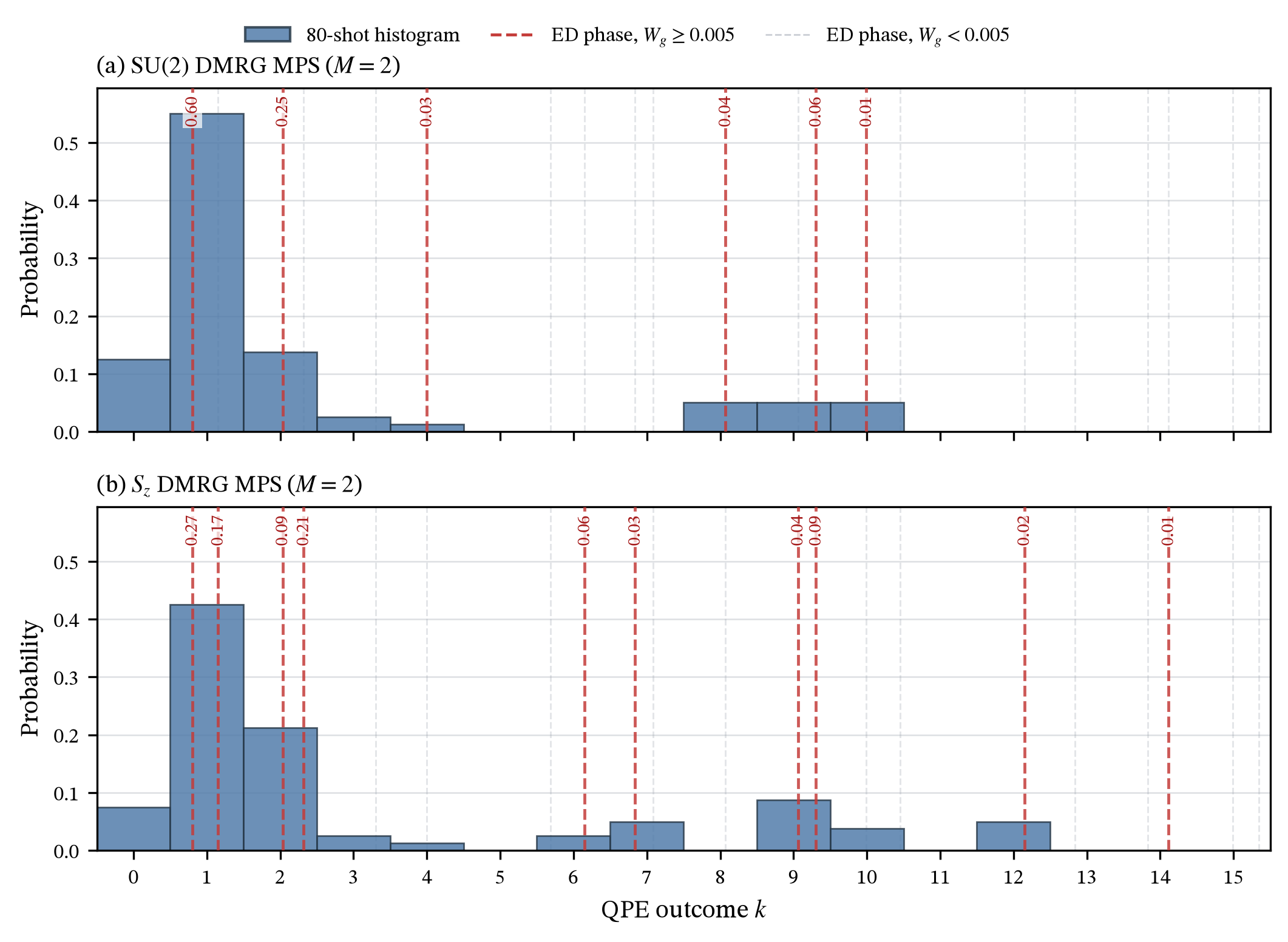}
    \caption{Sine-window QPE emulation for the $M=2$ SU(2) (a) and $S_z$ (b) DMRG states of the half-filled $2\times2$ Hubbard model at $U/t=8$. Histograms show one reproducible 80-shot realization with $t_q=4$.  Dashed vertical lines mark the exact ED phases. Red labels give the basis-invariant total weight $W_g=\sum_{n\in g}|c_n|^2$ of each degenerate energy manifold when $W_g\geq0.005$; lower-weight manifolds are shown as faint gray guides and are not labeled.}
    \label{fig:qpe_su2_sz_m2}
\end{figure}

The SU(2)-adapted MPS state shows $59.94\%$ of overlap weight on the exact ground state, more than twice the $27.12\%$ obtained from the $S_z$-adapted MPS state at the same nominal $M$.  It also assigns $84.77\%$ of its weight to negative-energy eigenstates and produces a visibly more concentrated low-$k$ QPE signal.  In the displayed finite-shot realization, 65 of 80 SU(2) samples lie in bins $k\leq2$, compared with 57 of 80 for the $S_z$ state.  These counts are not direct fidelity estimators because multiple nearby eigenstate phases contribute to the same finite-resolution bins, but they expose the different spectral concentration of the two trial states.

The distinction is both symmetry-driven and variational. The SU(2)-adapted DMRG state is a singlet to numerical precision and therefore has vanishing overlap with eigenstates in spin-incompatible sectors. By contrast, imposing only $S_z=0$ allows contributions from multiple total-spin sectors as indicated by many additional red phase lines in \cref{fig:qpe_su2_sz_m2}(b). MPS assigns a weight of approximately \(0.17\) to the lowest-energy excited state, which is a triplet. With four phase qubits, the corresponding eigenphase lies too close to the ground-state phase to be resolved as a distinct peak. Its contribution therefore merges with the ground-state QPE feature, reducing the accuracy with which the ground-state energy can be identified at this phase resolution. Consequently, in this strongly truncated example, SU(2) adaptation both increases the ground-state success probability of QPE and suppresses spectral leakage into symmetry-incompatible states.

\end{appendices}

\end{document}